\documentclass[oldversion]{aa}  
\usepackage{graphicx}
\usepackage{txfonts}
\usepackage{natbib}
\usepackage{aalongtable}

\newcommand\ud{\textrm{d}}

\def\0{\phantom0}

\def\obj{QSO~2237$+$0305}

\begin{document}

\title{Microlensing variability in the gravitationally lensed quasar \\    
\vspace*{1mm}  \obj\ $\equiv$ the Einstein Cross
 \thanks{Based on observations made with ESO Telescopes at the Paranal Observatory 
 under program ID 
 073.B-0243,
 074.B-0270, 
 075.B-0350, 
 076.B-0197,
 177.B-0615.}
 }
\subtitle{II. Energy profile of the accretion disk}

\author{
A. Eigenbrod\inst{1} \and F. Courbin\inst{1} 
\and G. Meylan\inst{1} \and E. Agol\inst{2} 
\and T. Anguita\inst{3} \and R. W. Schmidt\inst{3} \and J. Wambsganss\inst{3}}


\institute{
Laboratoire d'Astrophysique, Ecole Polytechnique F\'ed\'erale
de Lausanne (EPFL), Observatoire de Sauverny, 1290 Versoix, Switzerland
\and
Astronomy Department, University of Washington, 
Box 351580, Seattle, WA 98195, USA
\and
Astronomisches Rechen-Institut, Zentrum f\"ur Astronomie der Universit\"at Heidelberg, 
M\"onchhofstrasse 12-14, 69120 Heidelberg, Germany
}

\date{Received ... ; accepted ...}
\authorrunning{A. Eigenbrod et al.}
\titlerunning{Microlensing variability in the Einstein Cross}

\abstract{
We present the continuation of 
our long-term 
spectroscopic monitoring of 
the gravitationally lensed quasar \obj. We investigate  the chromatic
variations observed in the UV/optical continuum of both quasar images A and B, and compare 
them with numerical simulations to infer the energy profile of the
quasar accretion disk. Our procedure combines the microlensing ray-shooting technique 
with Bayesian analysis, and derives probability distributions for the source sizes as a 
function of wavelength. 
We find that the effective caustic crossing timescale  
is $4.0\pm 1.0$~months.
Using a robust prior on the effective transverse velocity,
we find that the source responsible for the UV/optical continuum has
an energy profile well reproduced by a power-law $R \propto \lambda^{\zeta}$
with $\zeta=1.2\pm0.3$, where $R$ is the source size responsible for the emission
at wavelength $\lambda$.
This is   the  first   accurate,   model-independent
determination of  the energy profile  of a quasar accretion  disk on
such small scales.

}


\maketitle

\section{Introduction}

Soon after quasars were discovered, it was suggested that they 
are powered by the accretion of gas on to supermassive black 
holes at the centers of galaxies. 
Since the infalling matter has non-zero angular momentum, it
forms a disk spinning around the central black hole \citep{lynden_bell69}. 
Viscosity within the disk is thought to result 
in an outward transfer of angular momentum, 
thereby allowing the material to spiral gradually inward,
heating the disk and causing it to emit intensely over 
a wide range of wavelengths \citep{shields78}.  

Despite the numerous studies addressing the subject, there
are still very few direct observational constraints on
the spatial structure of quasar accretion disks. Quasars
are located at 
cosmological distances, and it is particularly
difficult to observe the inner regions of these objects.
Direct imaging of the inner milli-parsec of a quasar would require 
angular resolutions on the order of  micro- or even nano-arcseconds.
There is currently no instrument capable of  such performance.
Fortunately, nature has provided us with a cosmic magnifying 
glass in the form of gravitational lensing, which
helps to resolve the source, and to reveal 
the spatial structure of its most inner parts.

Our target is \obj. It was discovered  by \cite{huchra} during the Center for
Astrophysics Redshift Survey, and is  also known as ``Huchra's
lens'' or the ``Einstein Cross''. It
is probably the most studied gravitationally
lensed quasar. It consists of a $z_s=1.695$ source
gravitationally   lensed  into four images   arranged   in a cross-like
pattern  around the nucleus  of a $z_l=0.039$  barred Sab galaxy. The
average projected distance  of  the images  from  the  lens center  is
$0.9$\arcsec. 

Gravitational lensing is achromatic, in the sense that photons are deflected
the same way regardless of their wavelength. However, lensing is sensitive
to the source size, and quasar accretion disks are known to vary chromatically
from the center to the edge. As a consequence, microlensing of the disk by stars
located in the lensing galaxy will affect the quasar images independently, inducing
chromatic differences in the spectra of the images \citep{wambsganss}. These 
differences are directly linked to the energy profile of the disk, i.e. the 
scaling between the wavelength and the corresponding size of the source emitting at 
that wavelength. 
As a consequence, microlensing-induced flux variations can be used to
constrain the energy profile of the quasar accretion disk.

\cite{rauch91} were the first to use this technique, and for the
Einstein Cross, they found that the  near-ultraviolet emission regions 
were smaller than expected for thermal
emission from an accretion disk. However, more modern work in this system and other
gravitational lenses have generally reached the opposite conclusion.
For instance, \cite{pooley07} finds that the
near-ultraviolet emission regions of ten lensed quasars 
are a factor of $3-30$ larger 
than the size predicted by simple accretion disk models 
to produce the observed optical flux.
In a more quantitative analysis, \cite{morgan07} 
use the microlensing variability observed for nine 
gravitationally lensed quasars to show that the accretion 
disk size is consistent with the expectation from 
thin disk theory \citep{shakura73}. However, these sizes are
larger, by a factor of $\sim 3$, than the size needed to produce the
observed infrared flux by thermal radiation from a thin disk. 
%


The Einstein Cross is particularly well suited for microlensing
studies, because of the symmetric configuration of the lensed images, which
 results in very short time delays,
and because of the low redshift of   the lensing galaxy, which places the
lensed images  in the bulge of the lensing  galaxy, where the
probability of microlensing by stars is high. 
Because of this, the Einstein Cross has been intensely monitored by different projects
\cite[e.g.,][]{corrigan91,ostensen96,alcalde02,schmidt02}.
The project having the longest  duration
and the best sampling rate is the Optical Gravitational Lensing Experiment
(OGLE) \citep{wozniak,wozniak00,udalski06}. OGLE has monitored \obj\ since 1997, 
and delivers the most complete lightcurves for this system.

Unfortunately, these monitoring campaigns are usually conducted 
in one photometric band, and hence they can be used to constrain 
the source size, but not the energy profile, which requires at least two bands.
Multi-band monitoring is one approach, but even more effective is long-term 
spectrophotometric monitoring, as described in our first 
paper \citep[][paper I in the following]{eigenbrod08}. 
In this previous contribution, we describe our observations and data analysis, and
report significant flux variations
in the continuum and broad emission lines of the spectra of the four 
lensed images, indicating that both the continuum emitting region
 and the broad line region 
are microlensed.  

Quasars are known to vary intrinsically, and in order to extract the 
microlensing-induced flux variations, we need to remove the intrinsic
fluctuations from the lightcurves. In practice we do this by considering 
the difference between the light curves of two images. In the present
study we focus on the lightcurves of images A and B, because they are
the two images undergoing the strongest flux variations within
the time span of our observations. In
the case of  \obj, we can neglect  the time delays between the images,
because they are  expected to  be  on the order of   one day or   less
\citep{schneider88,  rix92,wambsganss94}.  We also know  that the  observed
flux $F$ of a lensed quasar image is the product of the unlensed flux
$F_0$ of the source, the   extinction $e^{-\tau}$,  and the macro  and
micro-magnification $\mu_{macro}$   and $\mu_{micro}$    
$$ 
F=\mu_{micro} ~  \mu_{macro}  ~e^{-\tau}~  F_0 ~\textrm{.}
$$  
As observed in paper I, image A is affected by long-term (more than 5 years) 
microlensing. Hence the micro-magnification $\mu_{micro}$ 
has two components over the time span of our observation: one constant 
long-term magnification $\tilde{\mu}_{micro}$,
and one variable short-term $\mu^{\prime}_{micro}(t)$ 
$$
\mu_{micro}(t)= \mu^{\prime}_{micro}(t) ~\tilde{\mu}_{micro} ~\textrm{.}
$$
We shall mention that this separation in long and short-term microlensing
is purely empirical, and only reflects the impossibility of retrieving the
intrinsic source flux.
The difference in magnitude $\Delta m$ between image A and B is 
given by $\Delta m = -2.5 \log(F_A/F_B)$.
As observed in paper I, the extinction remains constant in 
time in all four images of the Einstein Cross.   The
macro-magnification given  in Table~\ref{tabrob} results in a constant
magnitude  difference on the order of  $-0.1$~mag between images A and
B.  As a consequence, the time variability that we observe in $\Delta m$
is only due to short-term microlensing   
$$
\Delta m 
= -2.5 \log\left(\frac{\mu^{\prime}_{micro,A}}{\mu^{\prime}_{micro.B}}\right)
  + m_0
$$
where $m_0$ is a constant. Microlensing depends on the source size.
Smaller sources are more strongly affected by microlensing than larger
sources. We also know that bluer photons are emitted closer to
the center of a quasar than redder ones. Hence we expect stronger 
variations of $\Delta m$ at bluer wavelengths \citep{wambsganss}.

In this paper, we use this
chromatic behavior of microlensing to compare the
observed  variability of  $\Delta m$ at different wavelengths
with numerical simulations.
Using Bayesian analysis similar to \cite{kochanek04} and \cite{anguita08}, we
derive the probability distributions for the size of the source emitting
at a given wavelength. This eventually defines the energy profile of the quasar
accretion disk.

In the following, we consider a flat cosmology with 
$(\Omega_m,\,\Omega_\Lambda)=(0.3,\,0.7)$  and $H_0=70$~km~s$^{-1}$~Mpc$^{-1}$.

\begin{table}[t]
\caption[]{Journal of the second part of our 
spectroscopic monitoring of \obj.
The Julian dates are given in HJD-2450000.}
\label{journal}
\begin{flushleft}
\begin{tabular}{clcccc}
\hline 
\hline 
ID & Civil Date & HJD & Mask & Seeing $[\arcsec]$ & Airmass \\
\hline  					       
32 & 10$-$07$-$2007& 4292 & 1 & 0.63 & 1.153 \\
   &               &      & 2 & 0.55 & 1.132 \\
33 & 15$-$07$-$2007& 4297 & 1 & 0.57 & 1.158 \\
   &               &      & 2 & 0.54 & 1.220 \\
34 & 25$-$07$-$2007& 4307 & 1 & 0.79 & 1.231 \\
   &               &      & 2 & 0.83 & 1.161 \\
35 & 03$-$08$-$2007& 4316 & 1 & 0.68 & 1.412 \\
   &               &      & 2 & 0.68 & 1.278 \\
36 & 27$-$08$-$2007& 4340 & 1 & 0.88 & 1.133 \\
   &               &      & 2 & 0.89 & 1.143 \\
37 & 06$-$09$-$2007& 4350 & 1 & 0.67 & 1.396 \\
   &               &      & 2 & 0.72 & 1.252 \\
38 & 20$-$09$-$2007& 4364 & 1 & 0.73 & 1.230 \\
   &               &      & 2 & 0.83 & 1.167 \\
39 & 23$-$09$-$2007& 4367 & 1 & 1.23 & 1.530 \\
   &               &      & 2 & 1.10 & 1.344 \\
40 & 05$-$10$-$2007& 4379 & 1 & 0.59 & 1.153 \\  
   &               &      & 2 & 0.55 & 1.132 \\  
41 & 10$-$10$-$2007& 4384 & 1 & 0.76 & 1.283 \\       
   &               &      & 2 & 0.64 & 1.195 \\       
42 & 15$-$11$-$2007& 4420 & 1 & 1.07 & 1.189 \\       
   & 16$-$11$-$2007& 4421 & 2 & 0.80 & 1.148 \\       
43 & 01$-$12$-$2007& 4436 & 1 & 1.00 & 1.502 \\       
   &               &      & 2 & 0.82 & 1.318 \\       
\hline  					        
\end{tabular}					        
\end{flushleft} 				        
\end{table}	 


\begin{table}
\begin{center}
\caption{ \label{bands} Wavelength intervals and central wavelength $\lambda_{c}$
(in \AA\ measured in the source frame)
of the six photometric bands. 
}
\begin{tabular}{c c c} 
\hline
Band&  $[\lambda_{min},\,\lambda_{max}]$ & $\lambda_{c}$ \\
\hline      
1 & [1500,\,1750] & 1625 \\	
2 & [1750,\,2000] & 1875 \\	
3 & [2000,\,2250] & 2125 \\	
4 & [2250,\,2500] & 2375 \\	
5 & [2500,\,2750] & 2625 \\	
6 & [2750,\,3000] & 2875 \\	
\hline
\end{tabular}
\end{center}
\end{table}

\section{Observations}

We  use two different  data sets for this study.  The first one is the
well      sampled      V-band      lightcurves      of          the
OGLE\footnote{http://www.astrouw.edu.pl/$\sim$ogle/ogle3/huchra.html}
project  \citep{udalski06}, from   which we  select  the  data between
Julian days 2453126 (April 30, 2004)  and 
2454439 (December 4, 2007), and which  comprises 352 data points.  
We bin the data points separate by less than one day, and are left with
181 points for which we recompute the errors as described in section~\ref{micro_ogle}.

The second data set is our deep spectrophotometric monitoring 
obtained with the FORS1 spectrograph 
mounted on the Very Large Telescope (VLT) of the European Southern Observatory (ESO). 
In paper I, we presented the
spectra of  all   four quasar images on  31   different epochs between
October 2004 and December 2006. We have now completed one more year 
of monitoring, and  a total of  43 different epochs spanning more 
than three years until December 2007 are now available. 
The journal of the new observations is given in Table~\ref{journal}. 
We remind that we use two masks for the spectroscopic observations.
The mask 1 has a slit centered on quasar images A and D, 
while the mask 2 has a slit centered on B and C. 
In the present study, we are interested in the difference magnitude
$\Delta m$ between images A and B. Thus, we will only consider the epochs where both 
masks 1 and 2 (see paper I), and hence both quasar images
have been observed on the same night. 
We therefore drop the epochs with ID 7, 8, 10, and 30, which reduces the number of usable epochs to 39.
The reduction and further processing of the new data 
are done following the same procedure as described in paper I; 
we spatially   deconvolve  our
spectra  to remove the lensing galaxy,  and 
we decompose the spectra into a sum of broad emission lines, continuum and 
iron pseudo-continuum. We fit  the continuum with a power-law  
$$
f_{\nu} \propto \nu^{\alpha_{\nu}}
\qquad \Rightarrow \qquad
f_{\lambda}  = f_0  \left(\lambda/\lambda_0\right)^{-(2+\alpha_{\nu})}
$$  
where $\lambda_0=2000$~\AA\ in the source frame, 
and where $\alpha_{\nu}$ is  the commonly
used canonical   power index.
  
\begin{figure}[t!]
\begin{center}
\includegraphics[width=8.5cm]{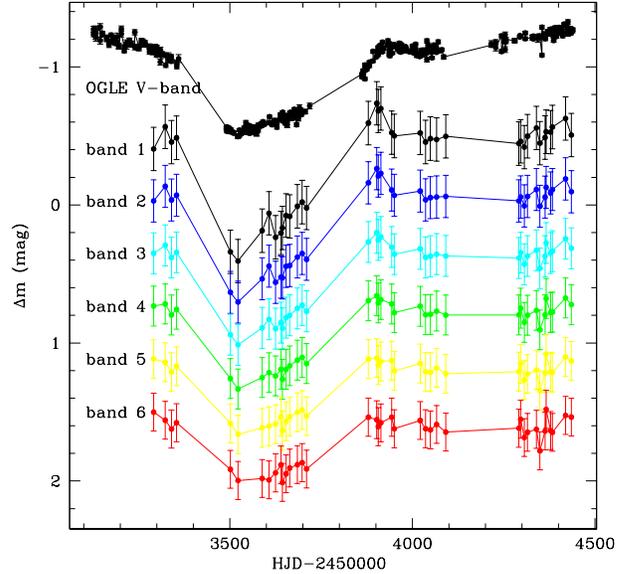}
\caption{The difference lightcurve $\Delta m$ between quasar images A and B
measured in different  wavelength bands. For clarity  we shift the curves along
the vertical direction.    The  binned OGLE observations with 181 data points
measured in   the  V-band   \citep{udalski06}.   From  our    39-epoch
spectrophotometric    monitoring, we compute the difference
lightcurves  of the continuum integrated in  the  six bands defined in
Table~\ref{bands}.}
\label{lightcurves}
\end{center}
\end{figure}

Based on this method,   we could identify strong
variations  in  the exponent $\alpha_{\nu}$   and intensity $f_0$  in the spectra
of quasar images A   and B.  This indicates that the  amplitude of  the
variations are  significantly higher in  the blue than in the
red part of the spectra.
In order to study in  more detail these chromatic fluctuations of the continuum, we
define  six wavelength  ranges,    that  we will  use   as  photometric
bands.  The bands all have the same width  and cover the whole wavelength
range between  1500 and 3000~\AA\ measured in  the source  frame,
see Table~\ref{bands}.   For each  of these bands we
compute the  $\Delta m$ lightcurve by integrating the continuum
power-law  in the corresponding  wavelength  range.
We give the corresponding AB magnitudes of the four quasar images 
in Table~\ref{mag_bands}. By subtracting
the magnitude of image B from that of image A, we obtain 
six 39-epoch lightcurves, as displayed in Fig.~\ref{lightcurves}.  
We immediately see that the  lightcurves  for the redder bands
(e.g. band \#6) vary less that the bluer ones (e.g. band \#1).

\begin{table}
\begin{center}
\caption{ \label{tabrob} Macro-lensing parameters for
images A and B from \cite{kochanek04} 
}
\begin{tabular}{c c c c}
\hline
Image&  $\kappa$   & $\gamma$ & $\mu_{macro} $ \\
\hline      
A & 0.394 & 0.395 & 4.735 \\
B & 0.375 & 0.390 & 4.192 \\
\hline
\end{tabular}
\end{center}
\end{table}

\section{Microlensing Simulations}
\label{micro_simul}

In order to simulate the microlensing effects, 
we have to choose a macro-model for the lensing galaxy. 
Many models have been proposed for the Einstein Cross
\citep[see for instance the summary in Table 2 of][]{wyithe02}.
We choose the  macro-model proposed by  \cite{kochanek04}. The surface
density $\kappa$, the shear $\gamma$, and the corresponding macro-magnifications 
$\mu_{macro}$ are given in  Table~\ref{tabrob}.
The convergence $\kappa$  is  usually separated into a  compact matter
distribution $\kappa_{\star}$   and  a  smooth    matter  distribution
$\kappa_c$ composed  of gas and dark  matter.  However, as  the quasar
images are located within  the bulge of the lensing  galaxy, we do not
expect a strong contribution  from the smooth matter \citep{kochanek07},
so we  neglect  $\kappa_c$, which amounts to assuming the entire 
surface mass density is in stars (i.e. $\kappa=\kappa_{\star}$).

The next step is to define the mass distribution of the microlenses.
Earlier investigations of microlensing 
have led to the conclusion that  
the magnification probability distribution 
is not very sensitive to the mass function of the microlenses
\citep[e.g.][]{wambsganss92, lewis95, wyithe01_320_21, congdon07},
but that  it  depends  on  the mean  microlens mass $\langle  M
\rangle$, which sets the   parameter scales, such as the  Einstein radius.
The determination of $\langle M \rangle$ is not trivial because of the
degeneracies  existing  between  $\langle M   \rangle$, the  effective
transverse   velocity, and the  source size. 
Several
studies succeeded in getting an estimate of $\langle M \rangle$, and
currently favor $\langle M    \rangle\simeq  0.1~M_{\odot}$
\citep[e.g.][]{wambsganss90b,lewis96}.   
We  will assume this  value for all
the microlenses distributed in the lens plane. 
This  value is also consistent  with the mean mass  found by the MACHO
collaboration who  studied  microlensing from objects  located towards
the bulge  of  the Milky Way \citep{alcock97a}. See also 
\cite{wyithe00_315_51}, who   give a  summary  of    the determinations   of
microlens masses in other galaxies.

The masses of the  microlenses sets the 
Einstein  radius $r_E$. For  \obj, the Einstein radius projects in the source plane as
$$
r_E=D_s~\sqrt{\frac{4 \mathrm{G} \langle M \rangle}{\mathrm{c}^2} \, \frac{D_{ls}}{D_s D_l}}
=5.77 \times 10^{16} ~ \langle M/0.1 M_{\odot}\rangle^{1/2}~\textrm{cm.}
$$
Having defined $\kappa$, $\gamma$ and the mass of the microlenses,
we can go one step further in our simulations, and start generating 
large magnification patterns
for  the two quasar images  A and B. We do  this using the inverse 
ray-shooting method \citep{wambsganssPhD,wambsganss90}. For each image, we
shoot approximately $10^{11}$ rays  from the observer through the lens
plane back   to the source plane,  where  the rays  are collected in a
$10,000$ by  $10,000$  pixels  array, corresponding to 
$100$  by  $100$ Einstein  radii. 



%


In order to study a sample of different source sizes, the  
magnification  patterns are convolved  with  a set of source
profiles. Microlensing-induced flux variations are
relatively insensitive to the source profile, but they depend
on its characteristic size. \cite{mortonson05}
showed that the half-light radius is the most important quantity 
for controlling the shape of microlensing light curves, whereas the
source profile is less important. For simplicity  we
choose a Gaussian profile for the surface brightness, and we
define the size of the source as the Full Width at Half Maximum (FWHM),
which is twice as large as the half-light radius, and 2.35 greater than
the Gaussian width $\sigma$.

The relative size of source with respect to the Einstein radius 
controls the smoothing of the magnification pattern.
Very large source sizes (i.e. larger than $4~r_E$) produce
magnification patterns that are so strongly smoothed, that they are
unable to account for 
the amplitude (i.e. higher than $1$~mag) of the flux variations seen in the 
OGLE lightcurves. Hence, as was shown in previous studies
\citep[e.g.][]{wyithe00_318_762,yonehara01},
we can safely rule out continuum source sizes larger than $4~r_E$.
On the other hand, very small source sizes 
(i.e. smaller than $0.01~r_E$) give very moderately smoothed
magnification patterns, which lead to  
sharp magnification events that are not observed in the OGLE lightcurves. 
The lower bound found in that way for the source size is $0.01~r_E$
\cite[e.g.][]{kochanek04, anguita08}.

In the present study, we vary the FWHM of  the Gaussian profile from  
1 to 400 pixels, which
corresponds  to 0.01 $r_E$ and 4.00  $r_E$, respectively. We consider
45 different source sizes between these two extreme values. 
They are 
  1,   2,   4,   6,   8,  10,  12,  14,  16,  18, 
 20,  22,  24,  26,  28,  30,  35,  40,  45,  50, 
 55,  60,  65,  70,  75,  80,  85,  90,  95, 100,
120, 140, 160, 180, 200, 220, 240, 260, 280, 300, 
320, 340, 360, 380, and 400 
pixels.

\section{Microlensing simulations fitting the OGLE data}
\label{micro_ogle}

Our simulations are conducted following the method described in
\cite{kochanek04}. This technique is based on a Bayesian analysis, that 
determines the probability distributions for the physical 
parameters of interest by comparing 
trial lightcurves with the observed data. Because we want to infer
probability distributions, we need to simulate
a large number of these trial lightcurves in order to obtain a
statistically significant sample. 
Thus, for each of our 45 source sizes,  we simulate 10,000
light curves for  both  quasar images A and B by tracing source trajectories
across the corresponding magnification patterns. 
We extract the pixel
counts  along the   positions   of  the  trajectory   using  bi-linear
interpolation and we convert  them   into magnitudes.  We
subtract the simulated  lightcurve of image B  from that  of image A,
and obtain a total of $4.5\times10^5$ difference lightcurves (10,000 for 45
different source  sizes). Eventually, we want to compare this large library of 
simulated difference lightcurves to the $\Delta m$ observed in the OGLE data, 
but before we do that, we specify some characteristics of the simulated 
source trajectories. 

The simulated lightcurves are obtained from source trajectories
across the magnification patterns, and these trajectories are characterized 
by parameters of two kinds: physical
and  trajectory parameters. The physical parameters 
are the local magnification tensor 
$\kappa$ and $\gamma$, the mass $\langle M \rangle$ of the microlenses,
the size $R_s$ of the source, the effective transverse velocity $V$ of
the source, and the magnitude offset $m_0$  between the two images due
to  a   combination  of   the  macro-magnification, 
long-term microlensing, and   differential
extinction between the images.
Because the scales of the magnification patterns are 
defined in terms of Einstein radii, our computational variables for the 
source size and velocity
are in fact the scaled source size
$\hat{R}_s=R_s/\langle M /0.1~M_{\odot} \rangle^{1/2}$ and the 
scaled transverse velocity $\hat{V}=V/\langle M /0.1~M_{\odot} \rangle^{1/2}$.
Following these definitions,
$\hat{R}_s$ is the relative size of the source with respect to 
the Einstein radius, and $\hat{V}$ is the velocity with which the source
is moving across the magnification pattern.

%

Each source trajectory is also defined by the trajectory
parameters  
$(\theta, \mathbf{x}_{0,A}, \mathbf{x}_{0,B})$, where 
$\theta$ is the direction angle, $\mathbf{x}_{0,A}$ are the coordinates of the 
starting point in the magnification pattern  of image A, and
$\mathbf{x}_{0,B}$ in the pattern of image B.
The trajectory is constrained to have identical directions $\theta$ and velocities $V$
across the patterns of  images A and B. 
The direction is set  to be the same
in  both patterns because the shear  direction between images A and B
is approximately the same \cite[]{witt94}, and because we assume 
the motion of the source to be primarily due to  the bulk motion of the
lensing galaxy rather than to the individual motions of the stars.
\cite{kundic93} and \cite{wambsganss95} show that the velocity dispersion of the stars
can statistically be interpreted as a bulk velocity 
artificially increased by an efficiency factor $a \simeq 1.3$.


For each particular choice of the parameters, that we write 
$p=(\kappa, \gamma, \langle M \rangle, \hat{R}_s, 
\hat{V}, m_0,\theta, \mathbf{x}_{0,A}, \mathbf{x}_{0,B})$,
we get one simulated difference lightcurve $\Delta
m^{\prime}_k(p)$, that  can   be compared to
the observed  data $D=\Delta  m_k$ by measuring  the
goodness of fit with a $\chi^2$ statistic 
$$
\chi_{OGLE}^2(p) = \sum_{k=1}^{n_{obs}} 
\left( \frac{\Delta m_{k} - \Delta m_{k}^{\prime}(p)}{\sigma_{k}} \right)^2
$$  
where $\sigma_k$ are the uncertainties of the OGLE data, and 
$n_{obs}=181$ is the number of the binned OGLE data points.
We determine the $\sigma_k$ from the photometric errors $\sigma_{OGLE}$
given by the OGLE project, from the standard deviation
between the binned points $\sigma_{bin}$, and from the systematic error
of the OGLE data $\sigma_{sys}$,
with $\sigma_k^2=\sigma_{OGLE}^2+\sigma_{bin}^2+\sigma_{sys}^2$.
We estimate the systematic error $\sigma_{sys}$  by carrying
out a polynomial fit of the OGLE difference lightcurve $\Delta  m_k$, using 
a polynomial of high enough order that the residuals are uncorrelated, 
i.e. such that the auto-correlation function of the residuals reduces 
to less than 0.5 within a separation of a few data points. We find that a polynomial 
on order 7 is sufficient, and $\sigma_{sys}=0.03$~mag.

By varying   the  parameters  we   construct   our  library  of $4.5\times10^5$
lightcurves. This produces an ensemble of  models, which given the data
and using Bayesian analysis (see Section~\ref{bayes})
can infer the probability distributions for the parameters.
The size  of the magnification patterns  is approximately one hundred
times greater than the time
scale of the observations multiplied by the effective velocity. 
Hence the available parameter space is huge, and it is relatively 
easy to find  good fits to  the  data. 

In our simulations we fix the values of 
$\kappa$, $\gamma$, and $\langle M \rangle$, and are left 
with the following set of variable parameters 
$p=(\hat{R}_s, \hat{V}, m_0,\theta, \mathbf{x}_{0,A}, \mathbf{x}_{0,B})$. The
physical parameters that we want to investigate are
the source size  and the effective transverse velocity. 
The other parameters, i.e. $(m_0, \theta, \mathbf{x}_{0,A}, \mathbf{x}_{0,B})$,
have no direct physical implications, and can be
chosen arbitrarily. They are therefore called nuisance parameters.
The influence of these parameters on the inferred probability distributions
of $\hat{R}_s$ and $\hat{V}$ vanishes when using Bayesian analysis
and a sufficiently large library of simulated lightcurves, as 
described in Section~\ref{bayes}.

\begin{figure}[t!]
\begin{center}
\includegraphics[width=8.5cm]{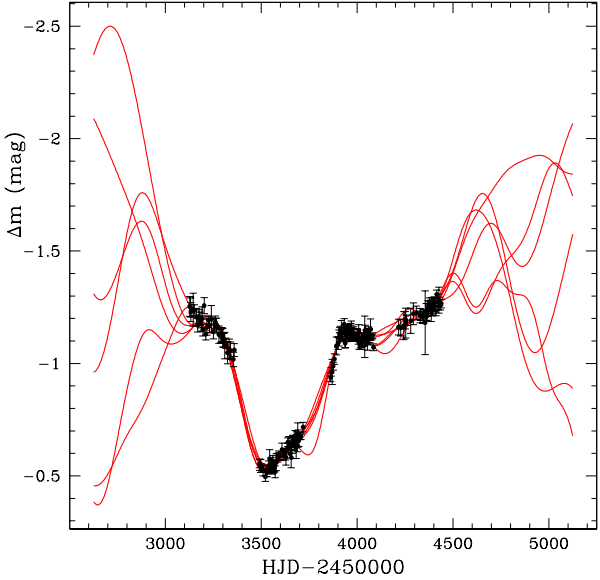}
\caption{Examples of six simulated lightcurves fitting the observed OGLE difference
lightcurve $\Delta m$ between quasar images A and B.}
\label{ex_lc}
\end{center}
\end{figure}

A first guess for the nuisance parameters  $(m_0, \theta, \mathbf{x}_{0,A},
\mathbf{x}_{0,B})$ is chosen randomly following a uniform
distribution. In the Bayesian analysis, lightcurves with low
$\chi_{OGLE}^2$ values have relatively higher weights, and 
contribute more to the final probability distributions. 
An effective and fast method consists in searching for
trajectories with low $\chi_{OGLE}^2$ values.
To do this and hence minimize the necessary computing time, we follow the procedure
described in \cite{anguita08}, and optimize the seven  parameters 
$(\hat{V}, m_0, \theta, \mathbf{x}_{0,A},\mathbf{x}_{0,B})$
with a
$\chi^2$-based minimization algorithm  using a   Levenberg-Marquardt  
least squares   routine  in
\textsc{idl}\footnote{http://cow.physics.wisc.edu/$\sim$craigm/idl/}. We
verify that  this minimization technique still  samples  uniformly
the whole magnification pattern. Finally we obtain a trajectory  library
containing  $4.5\times10^5$ trajectories   (10,000  trajectories  for 45
different source  sizes) fitting the OGLE data.
All the source trajectories have a reduced $\chi_{OGLE}^2/n_{dof} < 10$, where
$n_{dof}=345$ is the number of degrees of freedom, i.e. the number of observations
minus the number of fitted parameters (here 7).
Six examples of the simulated lightcurves are given in Fig.~\ref{ex_lc}.
 

\section{Bayesian analysis}
\label{bayes}

In the previous section, we have described how we build our  large library of trial lightcurves.
We will now use these lightcurves to determine the probability distributions 
for the parameters $\hat{R}_s$ and $\hat{V}$ based on Bayesian analysis as described by \cite{kochanek04}.
Following  Bayes' theorem the probability of the parameters
$p=(\kappa, \gamma, \langle M \rangle, \hat{R}_s, 
\hat{V}, m_0,\theta, \mathbf{x}_{0,A}, \mathbf{x}_{0,B})$, 
given the data $D=\Delta  m_{k}$, is
$$
P(p | D) 
= \frac{P(D | p)\, P(p)}{P(D)}
= \frac{L(D | p)}{N}\, P(p)
$$
where $P(p)$ is the prior, $L(D | p)$ is the likelihood and $N$ is
a normalization constant. 

\begin{figure*}[t!]
\begin{center}
\includegraphics[width=6cm]{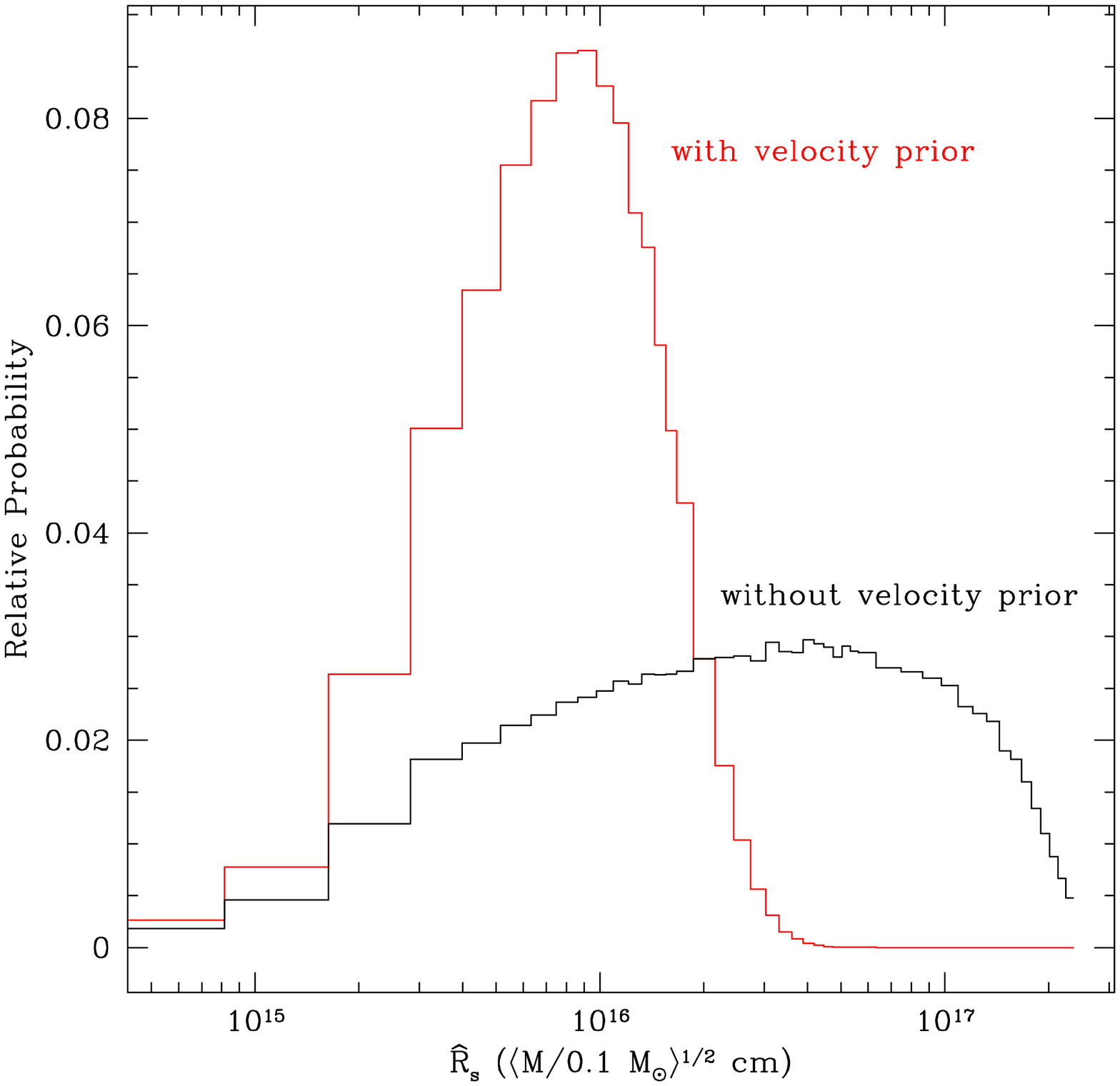}
\includegraphics[width=6cm]{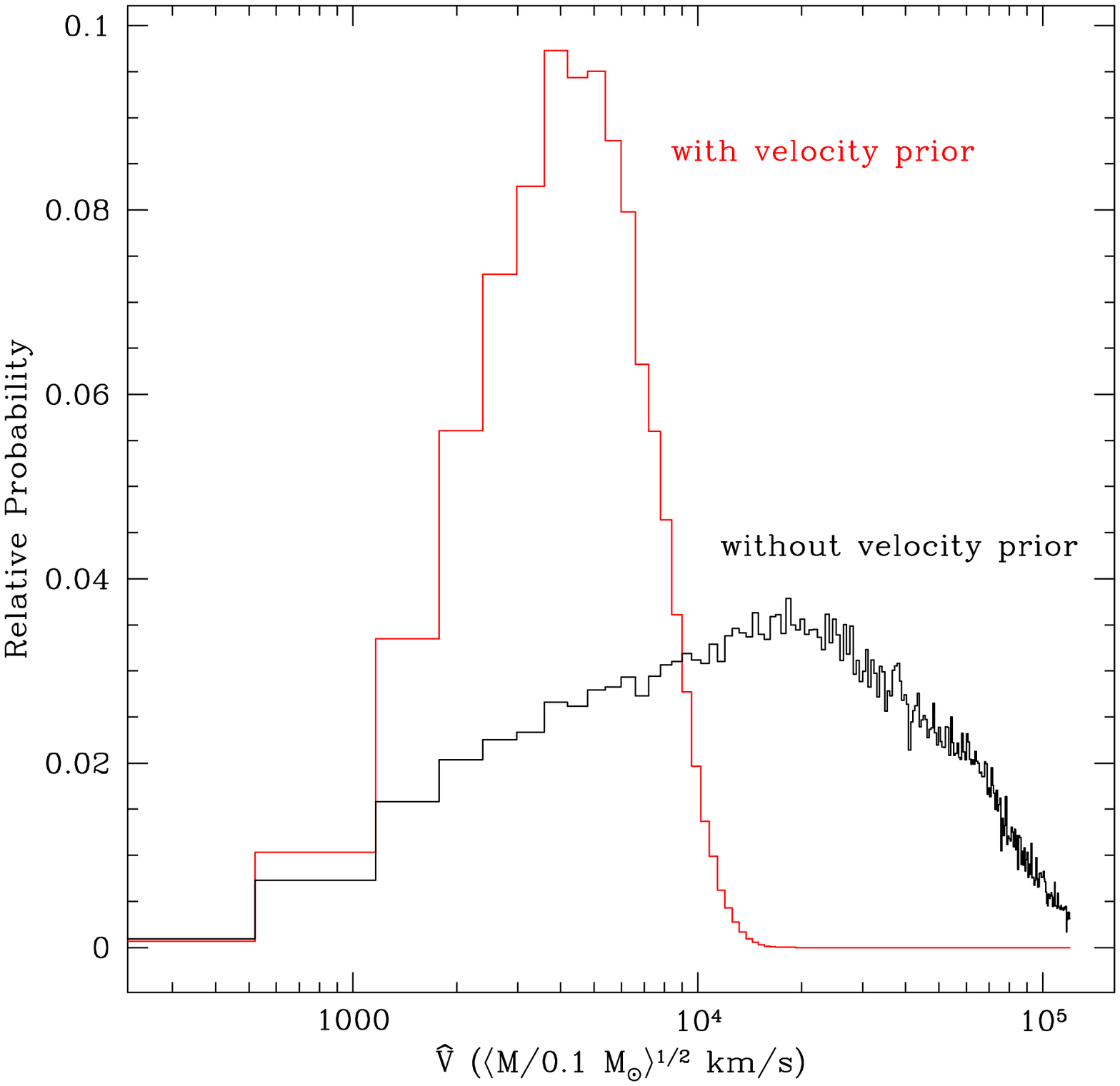}
\includegraphics[width=6cm]{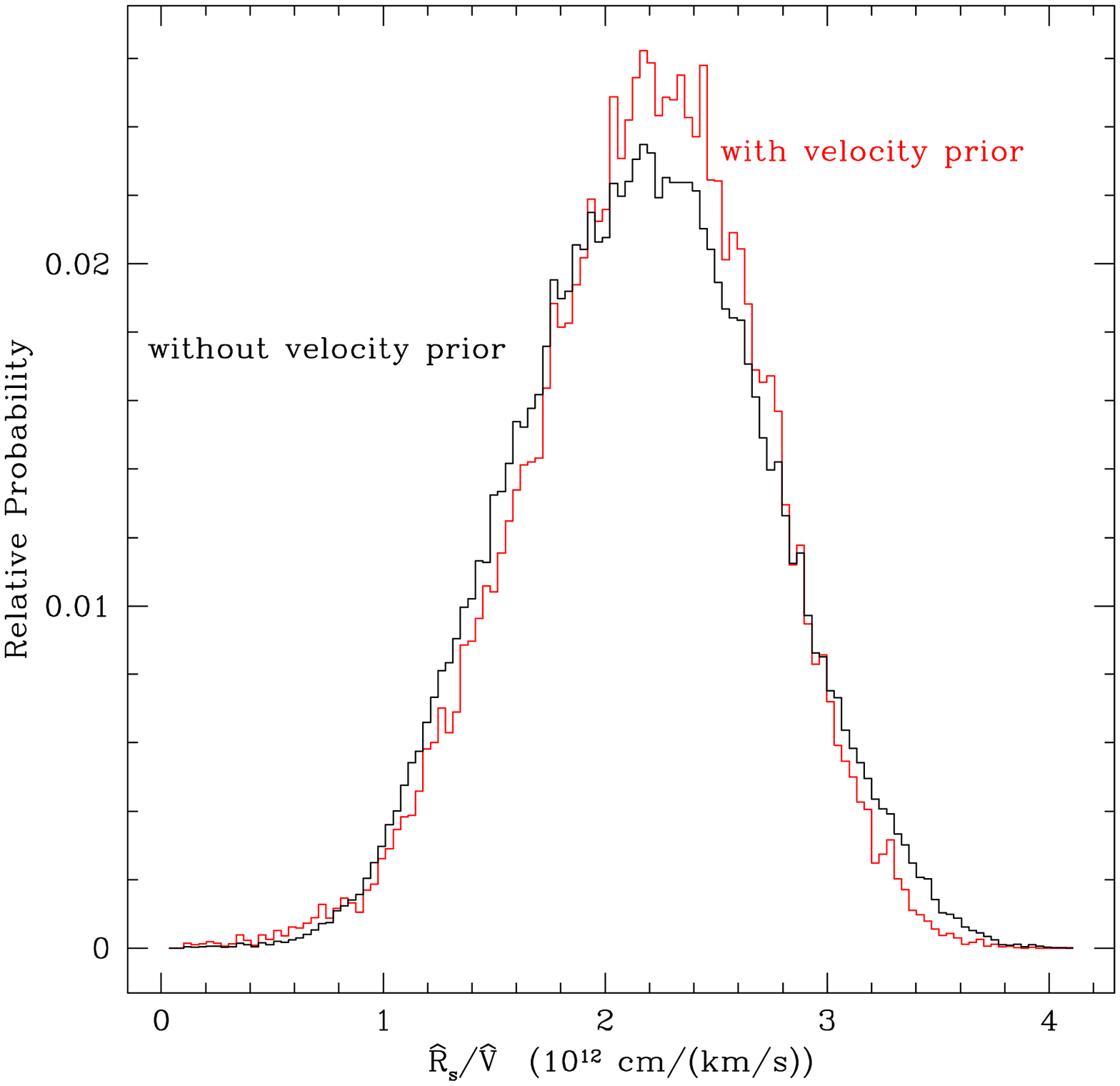}
\caption{\emph{Left :} The probability distributions for the FWHM $\hat{R}_s$ of the source, 
based on the OGLE data. We consider two cases, one with and another without the prior on 
the velocity.
\emph{Center :} Same as left panel, but for the effective transverse velocity $\hat{V}$.
\emph{Right :} The probability distributions for the ratio $\hat{R}_s/\hat{V}$.}
\label{distrib}
\end{center}
\end{figure*}

We do not have any prior information either on the trajectory parameters
$\theta, \mathbf{x}_{0,A},\mathbf{x}_{0,B}$  
or on the  magnitude offset $m_0$.  Therefore we choose
these  parameters from a  random and uniform  distribution. This means 
that the corresponding priors 
$P(m_0)$, $P(\theta)$, $P(\mathbf{x}_{0,A})$, and $P(\mathbf{x}_{0,B})$  
are constant. We also assume that the parameters are independent, i.e.
that 
$$
P(p)= 
P(\kappa) \, 
P(\gamma ) \, 
P(\langle M \rangle) \, 
P(\hat{R}_s) \, 
P(\hat{V}) \, 
P(m_0) \, 
P(\theta) \, 
P(\mathbf{x}_{0,A}) \, 
P(\mathbf{x}_{0,B})  ~\textrm{.}
$$
We define the relative likelihoods of the parameters $p$  based on the 
$\chi^2$ statistics. Usually, this is done following the standard approach for ensemble analysis
\citep[e.g.,][]{sambridge99}, which uses the  maximum likelihood estimator  
$$  
L(D | p)=\exp\left(-\frac{1}{2}\,\chi^2(p)\right)  ~\textrm{.}
$$ 
However, as noted by \cite{kochanek04}, using this standard estimator for
the OGLE data works poorly because 
we are comparing the probabilities of completely different light curves rather 
than models related to each other by
continuous changes of parameters. 
To circumvent this, we use the likelihood
estimator proposed by \cite{kochanek04}
$$  
L(D | p)=\Gamma\left[\frac{n_{dof}-2}{2},\frac{\chi^2(p)}{2}\right]
$$ 
where $\Gamma$ is an incomplete Gamma function.
Finally, all Bayesian parameter  estimates have to be normalized by the 
requirement that the total probability is unity, i.e. 
$\int P(p | D) ~\ud p =1$ and hence
$$
N=\int L(D |p) \, P(p)~\ud p ~\textrm{.}
$$
In practice we sum   the probabilities for    our random sampling  of
trajectories, which  is  equivalent to  using Monte  Carlo integration
methods to   compute the integral  over the  ensemble of  all possible
trajectories.
The sum over the random trajectories will converge to the true
integral provided we consider a sufficiently large number of 
trajectories. The probability
of a trajectory $j$ (defined by the set of parameters $p_j$)
given the data $D$ is 
\begin{equation}
\label{prob}
P(p_j|D)
=\frac{L(D | p_j) \, P(\hat{R}_{s,j}) \, P(\hat{V}_j)}{
\sum_{j=1}^{n} L(D | p_j) \, P(\hat{R}_{s,j}) \,P(\hat{V}_j)}
\end{equation}
where $n$ is the total number of trajectories in the library, 
and $P(\hat{R}_{s,j})$ and $P(\hat{V}_j)$ are the priors on the source size and 
velocity, respectively.
The priors on the other parameters are the same for all trajectories, 
and cancel out in the fraction above. 
The final probability distribution of the parameters are
obtained by summing the probabilities $P(p_j|D)$
of all the trajectories $j$ of the library.

The prior $P(\hat{R}_{s,j})$ is required because  
we have not uniformly sampled  the parameter
space of the source size. We have to correct for this.
The prior $P(\hat{R}_{s,j})$ is related to the density of trajectories
as a function of $\hat{R}_{s,j}$ given our sampling of the source size in 45 
different bins. 
Each bin contains 10,000 trajectories. Therefore, small
bins have higher densities, and 
the corresponding $P(\hat{R}_{s,j})$ should be proportionally lower to compensate 
this effect, i.e. if the size of bin $b$ is $l_b$, and if $\hat{R}_{s,j}$ falls
in this bin, then $P(\hat{R}_{s,j})\propto l_b$. Thus each bin has the same 
density of trajectories.

As already mentioned, there is a strong, essentially linear,
correlation between $\hat{R}_s$
and $\hat{V}$, i.e.  $\hat{R_s} \propto \hat{V}^x$
where $x\simeq 1$ \citep{kochanek04}.
Furthermore, $\hat{R}_s$ and $\hat{V}$ are both defined as
functions of $\langle M\rangle$, which imply  degeneracies
between $\langle M \rangle$, $\hat{V}$, and $\hat{R}_s$.
However, since $R_s= \hat{R}_s \langle M/0.1 M_{\odot} \rangle^{1/2}$ and 
$\langle M \rangle \propto (V/\hat{V})^2$, 
the physical size of the source $R_s \propto V \,\hat{V}^{x-1}\simeq V$
depends essentially on our estimate of the physical velocity $V$, and avoids 
the degeneracies between $\langle M \rangle$, $\hat{V}$, and $\hat{R}_s$.

Estimates of $V$ can be obtained from the
observations of the motion of other galaxies or galaxy clusters \citep[e.g.][]{benson03},
which show that peculiar velocities are typically not higher than $1,500$~km/s. 
In order to give a lower weight to trajectories 
having much higher velocities, we have to introduce a prior on the velocity.
Previous studies have considered various priors 
\citep[e.g.][]{kochanek04,gilmerino05}. Some are more restrictive than others,
but most of them have in common that they favor transverse velocities 
of $\sim600$~km/s in the lens plane, which correspond to projected
transverse velocities of $\sim6,000$~km/s in the source plane 
of \obj\ \citep{kayser86}.

The effective transverse velocity is the result of the relative
motion between the source, the lens and the observer \citep{kayser86}, 
and is enhanced by a contribution from the velocity dispersion of 
the stars in the lensing galaxy \citep{kundic93,wambsganss95}. 
The problem is that we do not known the peculiar velocity of either
the source, or the lens. The best we can do is to consider 
probability distributions for these unknown velocities, based
on what we know from the peculiar motion of other galaxies
\citep[e.g.][]{mould93,benson03}.
\cite{kochanek04} shows that a good approximation for
the probability distribution
of the effective transverse velocity in the source-plane is
\begin{equation}
\label{prior}
P(V)
= \frac{V}{\bar{V}} ~
I_0\left[\frac{V \bar{V}}{\sigma^2}\right] ~
\exp\left( -\frac{V^2+\bar{V}^2}{2 \sigma^2} \right)
\end{equation}
where $I_0$ is a modified Bessel function, $\bar{V}\simeq  2500$~km/s 
is the mean velocity, 
and $\sigma\simeq 3500$~km/s is the total root mean square velocity obtained by
considering all the different velocities contributing to the transverse 
effective velocity.
In our analysis we assume  $\langle M\rangle=0.1~M_{\odot}$, and 
we define the prior on the scaled velocity $\hat{V}$ as $P(\hat{V})=P(V)$.

Using Bayesian analysis, and considering both cases
with and without velocity prior, we obtain the probability distributions
of the source size $\hat{R}_s$, the effective transverse velocity $\hat{V}$, and
the ratio $\hat{R}_s/\hat{V}$ plotted in Fig.~\ref{distrib}. The results are 
given in Table~\ref{result}. 


\begin{figure*}[t!]
\begin{center}
\includegraphics[width=8.5cm]{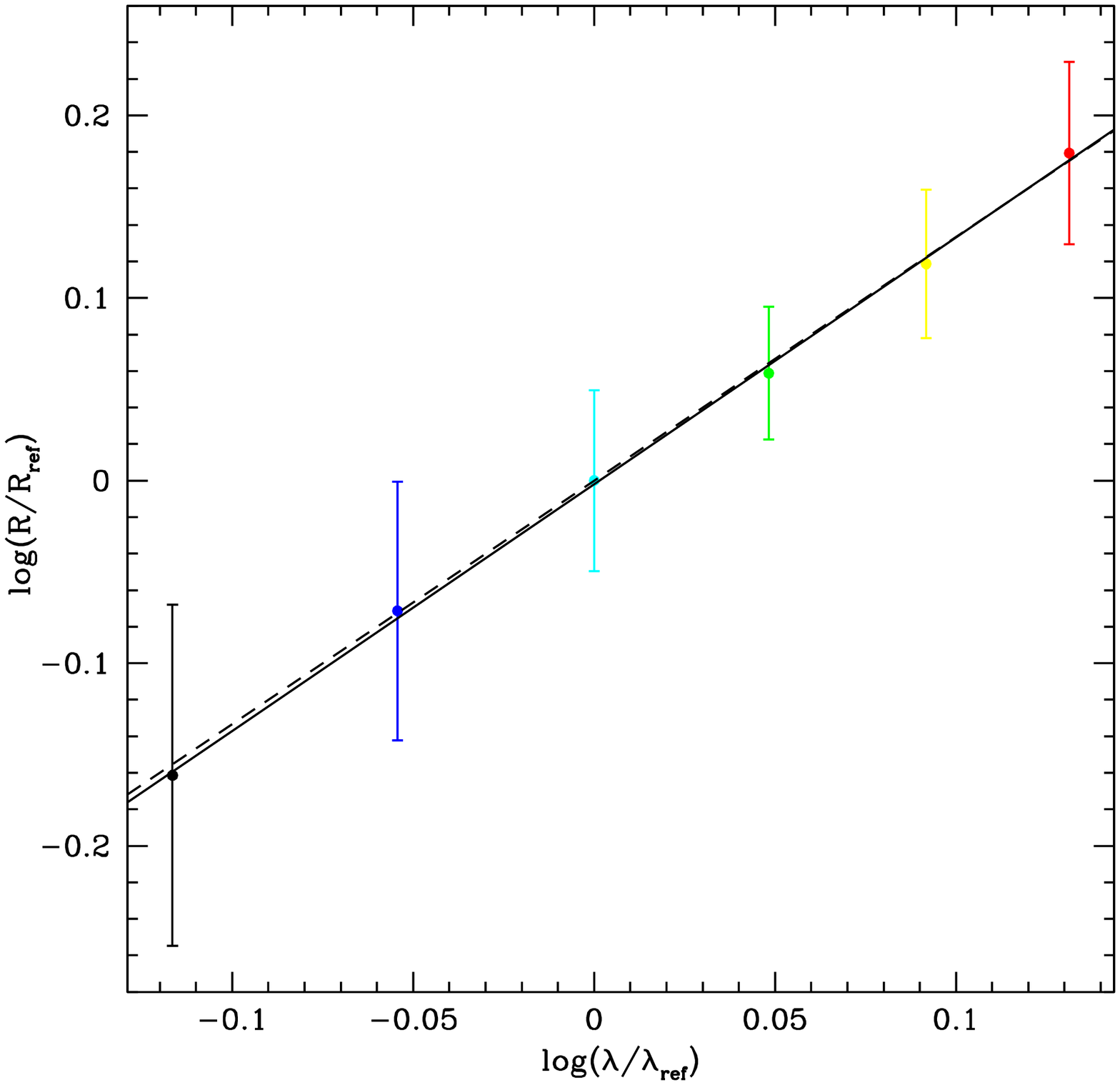}
\includegraphics[width=8.5cm]{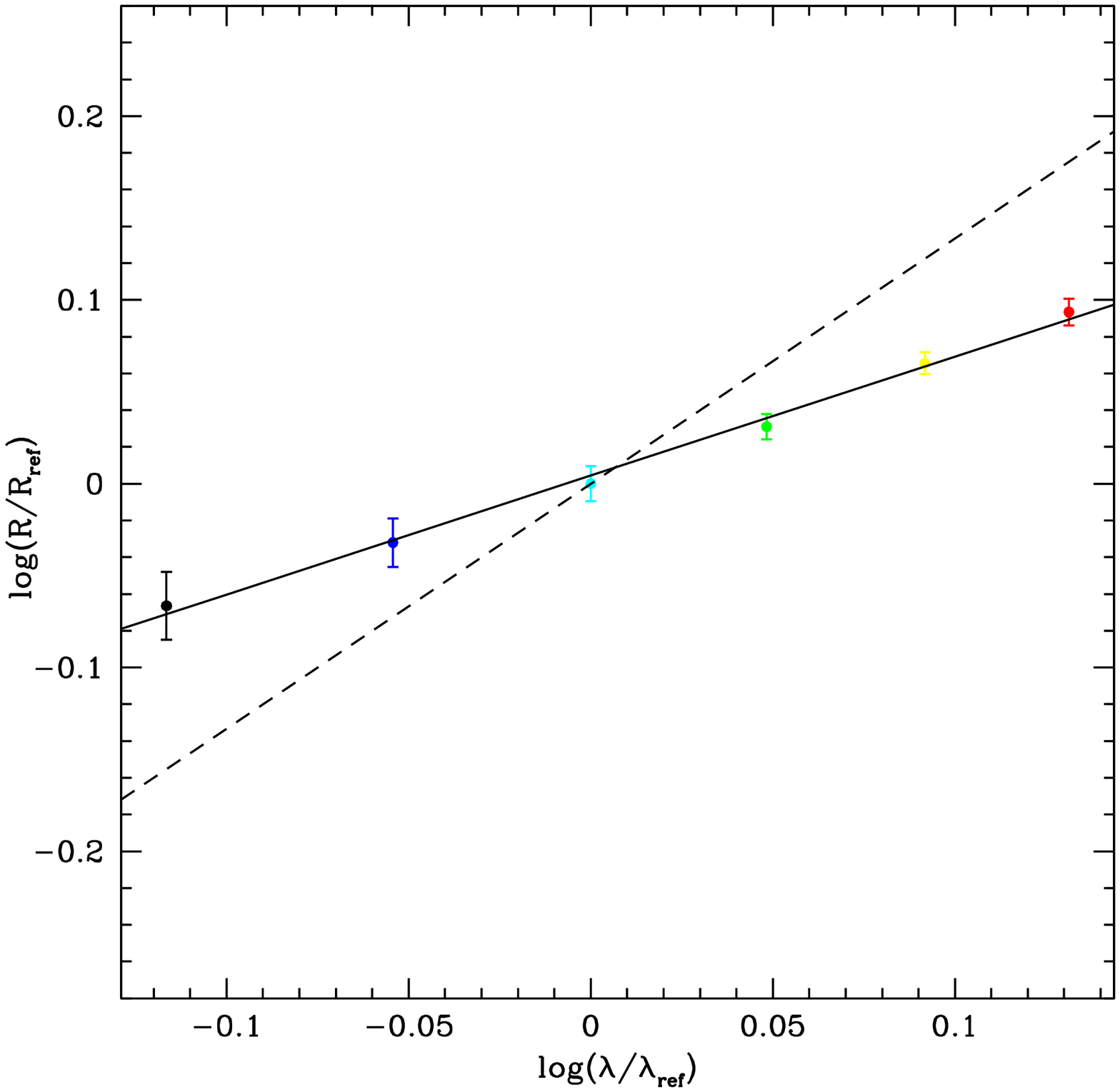}
\includegraphics[width=8.5cm]{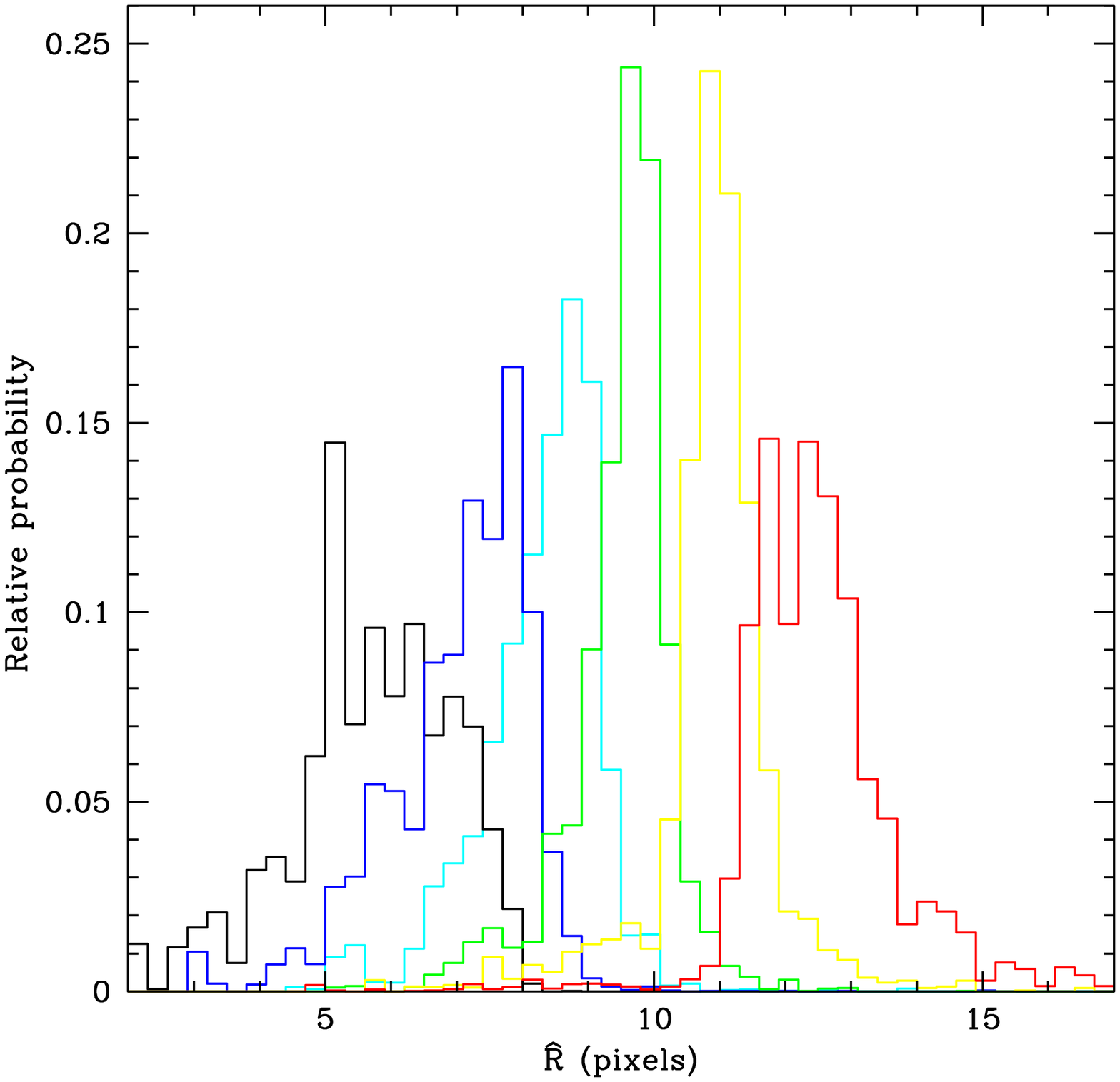}
\includegraphics[width=8.5cm]{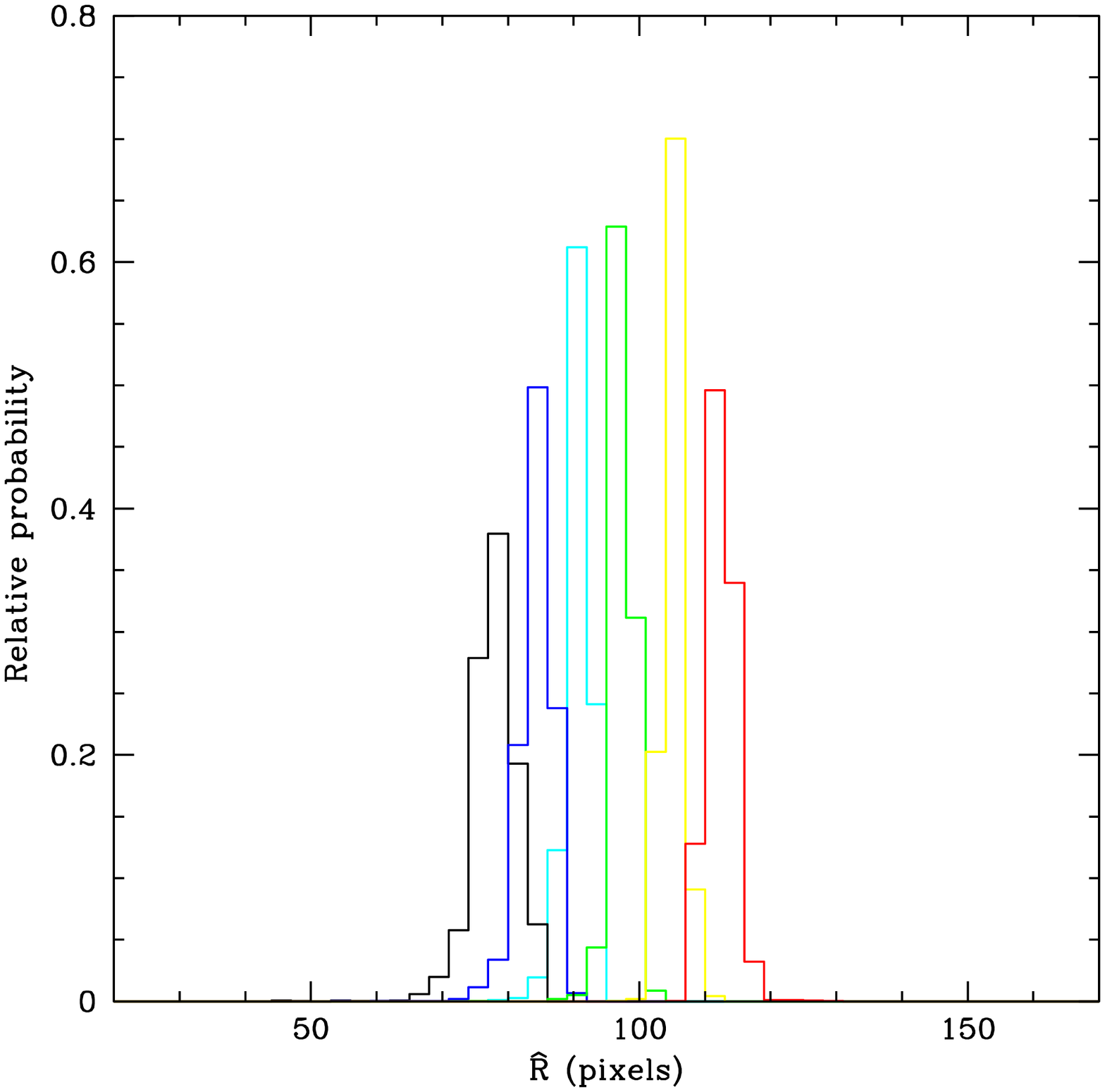}
\caption{\emph{Top :} The source FWHM ratio 
$R_{i}/R_{ref}=\hat{R}_{i}/\hat{R}_{ref}$ as a function of the wavelength ratio
$\lambda_i/\lambda_{ref}$ obtained from the trajectories in the library having 
$\hat{R}_s=0.1\,r_E=10$~pixels (left)
and $\hat{R}_s=1\,r_E=100$~pixels (right). 
The reference wavelength is $\lambda_{ref}=2125$~\AA\ measured in the source frame.
The error bars give the $1\sigma$ uncertainties. The solid line indicates the linear regression
across the points.
The dashed line shows the relation expected for
the standard $R\propto \lambda^{4/3}$ scaling.
\emph{Bottom :} The corresponding probability distributions of the scaled
source sizes $\hat{R}_i$
obtained in the six bands $i=1,2,..., 6$, and using the same color code as
in Fig.~\ref{lightcurves}.}
\label{R_lambda}
\end{center}
\end{figure*}

\begin{table}
\begin{center}
\caption{ \label{result} 
Results from the microlensing 
simulations and Bayesian analysis applied to the OGLE data.
}
\begin{tabular}{l }
\hline
with velocity prior  \\
$\qquad \hat{R}_s= \left(9.2^{+6.9}_{-5.8} \right) \times 10^{15}\, \langle M/0.1 M_{\odot}\rangle^{1/2}$~cm 
          $= \left(0.16^{+0.12}_{-0.10} \right) ~r_E$ \\
$\qquad \hat{V}  = \left(3.9^{+3.0}_{-1.8} \right) \times 10^{3} \, \langle M/0.1 M_{\odot}\rangle^{1/2}$~km/s\\
$\qquad \hat{R}_s/\hat{V}= \left(2.1 \pm 0.5  \right) \times 10^{12} $~cm/(km/s)\\
without velocity prior  \\
$\qquad \hat{R}_s= \left(4.0^{+7.5}_{-3.5} \right) \times 10^{16}\, \langle M/0.1 M_{\odot}\rangle^{1/2}$~cm 
          $= \left(0.69^{+1.30}_{-0.60} \right) ~r_E$ \\
$\qquad \hat{V}  = \left(1.8^{+2.8}_{-1.6} \right) \times 10^{4} \, \langle M/0.1 M_{\odot}\rangle^{1/2}$~km/s\\
$\qquad \hat{R}_s/\hat{V}= \left( 2.1 \pm 0.6 \right) \times 10^{12} $~cm/(km/s) \\
\hline
\end{tabular}
\end{center}
\end{table}


\section{Energy profile of the  quasar accretion disk}

Accretion disk models for quasars generally make the assumption that the
disk is optically thick and geometrically thin;  both assumptions are
required to cause efficient conversion of the gravitational potential energy
of accreting matter into radiation \citep{pringle72,shakura73,novikov73}.  
Alternative models in which mass is lost in a wind 
\citep{kuncic07}, the disk accretes at greater than the
Eddington limit and thus becomes geometrically thick, in which case
advection or convection can dominate the radial heat transport 
\citep[e.g.][]{abramowicz88}, or magnetic connections between the
black hole and disk modify the rate of energy dissipation \citep{agol00}.


The peak of the continuum emission in a typical quasar spectral energy
distribution is in the "Big Blue Bump", and can be fitted well with a spectrum which scales
as $F_\nu \propto \nu^{1/3} e^{h\nu/k T}$, with a characteristic
temperature $T = 7 \times 10^4$~K \citep{krolik98}.  
This is predicted by the standard thick
accretion disk model in which $T \propto R^{-3/4} (1-(R_{in}/R)^{1/2})^{1/4}$, the
functional form predicted by non-relativistic thin accretion disks with an
inner edge $R_{in}$ at which the viscous stress disappears 
\citep[relativity modifies this somewhat - ][]{krolik_book_98}.  
At large radius $T \propto R^{-3/4}$, and 
the model predicts that the
radiative flux $F$ (erg cm$^{-2}$ s$^{-1}$) at the surface of the disk at a
radius $R$ is proportional to the energy production rate at that radius,
resulting in $F \propto R^{-3}$.  Since we are assuming a
blackbody spectrum, $F \propto T^4$, and because the wavelength $\lambda$ of
the peak of the black body spectrum is proportional to $T^{-1}$, 
most of the radiation at a given radius $R$ comes from near a radius
$$
R 
\propto T^{-4/3} \propto \lambda^{4/3} ~\textrm{.}
$$
This defines the energy profile of the quasar, away from the inner edge of
the disk, which is identical to the form obtained by \cite{kochanek07}
from the standard thin accretion disk model of 
\cite{shakura73}.

However, given that we are trying to test the accretion disk model, we allow
the temperature to scale as a power-law with radius with arbitrary slope
$T \propto R^{-\zeta}$
\citep{rauch91}; this functional form is general enough to cover a
variety of different alternatives to the standard accretion disk model.

There exist numerous models for quasar accretion disks
in the literature. We can not consider all of them here, and we select
only a few  presenting an interesting variety.
One particular model is motivated by the fact that  
the theoretical optical/UV continuum of a standard thin disk is 
$F_{\nu} \propto \nu^{1/3}$, which is inconsistent with 
$F_{\nu} \propto \nu^{-1}$ observed in many quasars. Hence,
\cite{gaskell07} suggests that the observed quasar spectra can be 
reproduced by accretion disks with a temperature gradient of
$T \propto R^{-0.57}$ instead of $T \propto R^{-3/4}$. This implies 
$$
R 
\propto \lambda^{1.75} ~\textrm{.}
$$ 
Another model is proposed by  \cite{agol00}, 
and describes a disk which is powered by
the spin of the central black hole. This 
changes the temperature profile to $T \propto r^{-7/8}$, and
predicts 
$$
R \propto \lambda^{8/7} ~\textrm{.}
$$ 
The three models mentioned above all predict a 
power-law relation between the wavelength
and the corresponding source size 
$$
R\propto \lambda^{\zeta} 
$$
with different values for the index $\zeta$.
For two different radii
$R$ and $R_{ref}$ emitting at the wavelengths $\lambda$ and $\lambda_{ref}$
respectively, we have
$$
\frac{R}{R_{ref}}
=\left(\frac{\lambda}{\lambda_{ref}}\right)^{\zeta} 
\qquad \Rightarrow \qquad
\log\left(\frac{R}{R_{ref}}\right)
=\zeta~\log\left(\frac{\lambda}{\lambda_{ref}}\right) ~\textrm{.}
$$ 
To distinguish between these different accretion disk models, we 
determine in the next section, which power-law
is best compatible with our spectroscopic observations, and which 
of these models can be ruled out for \obj.

\begin{figure*}[t!]
\begin{center}
\includegraphics[width=8.5cm]{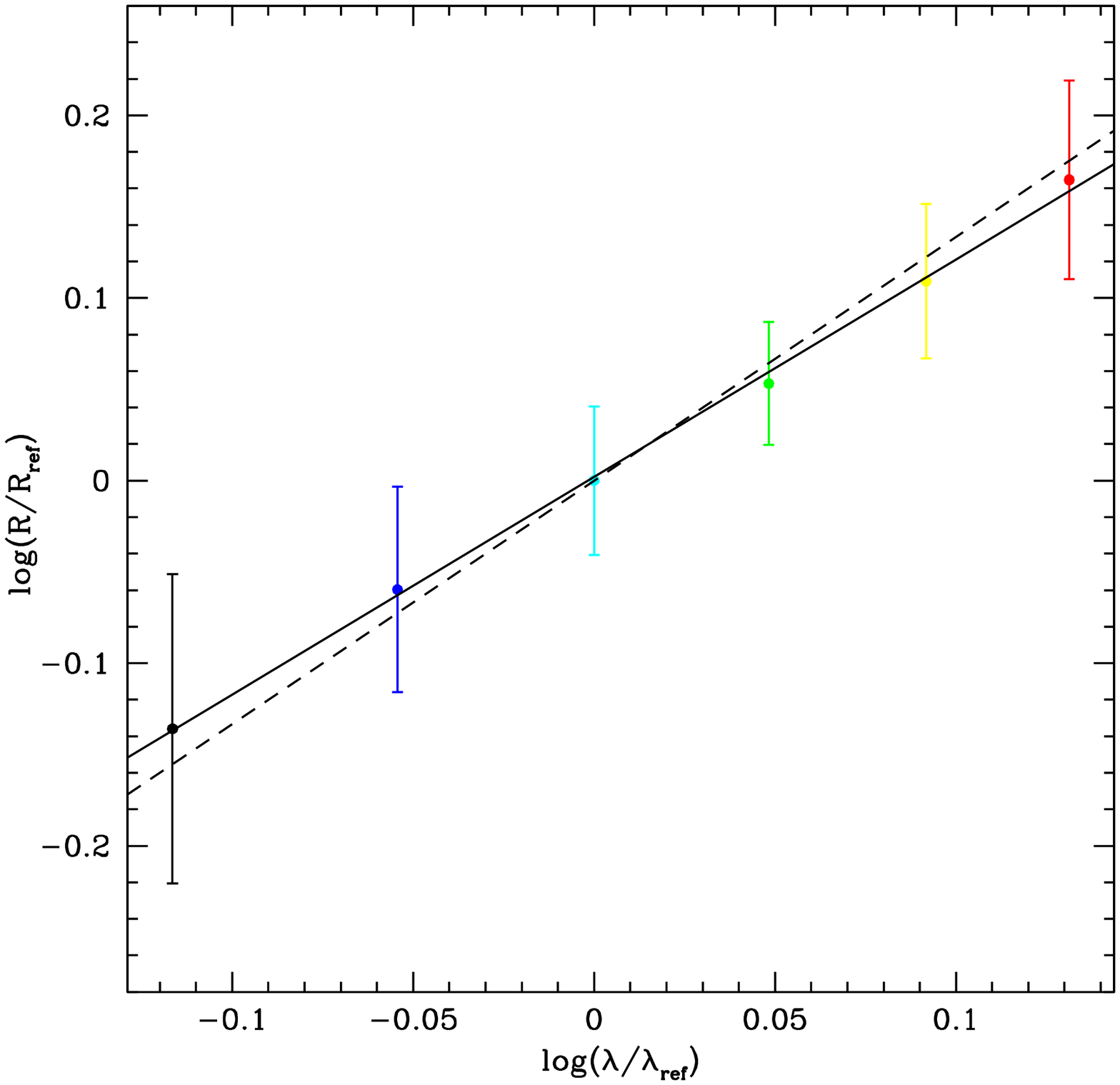}
\includegraphics[width=8.5cm]{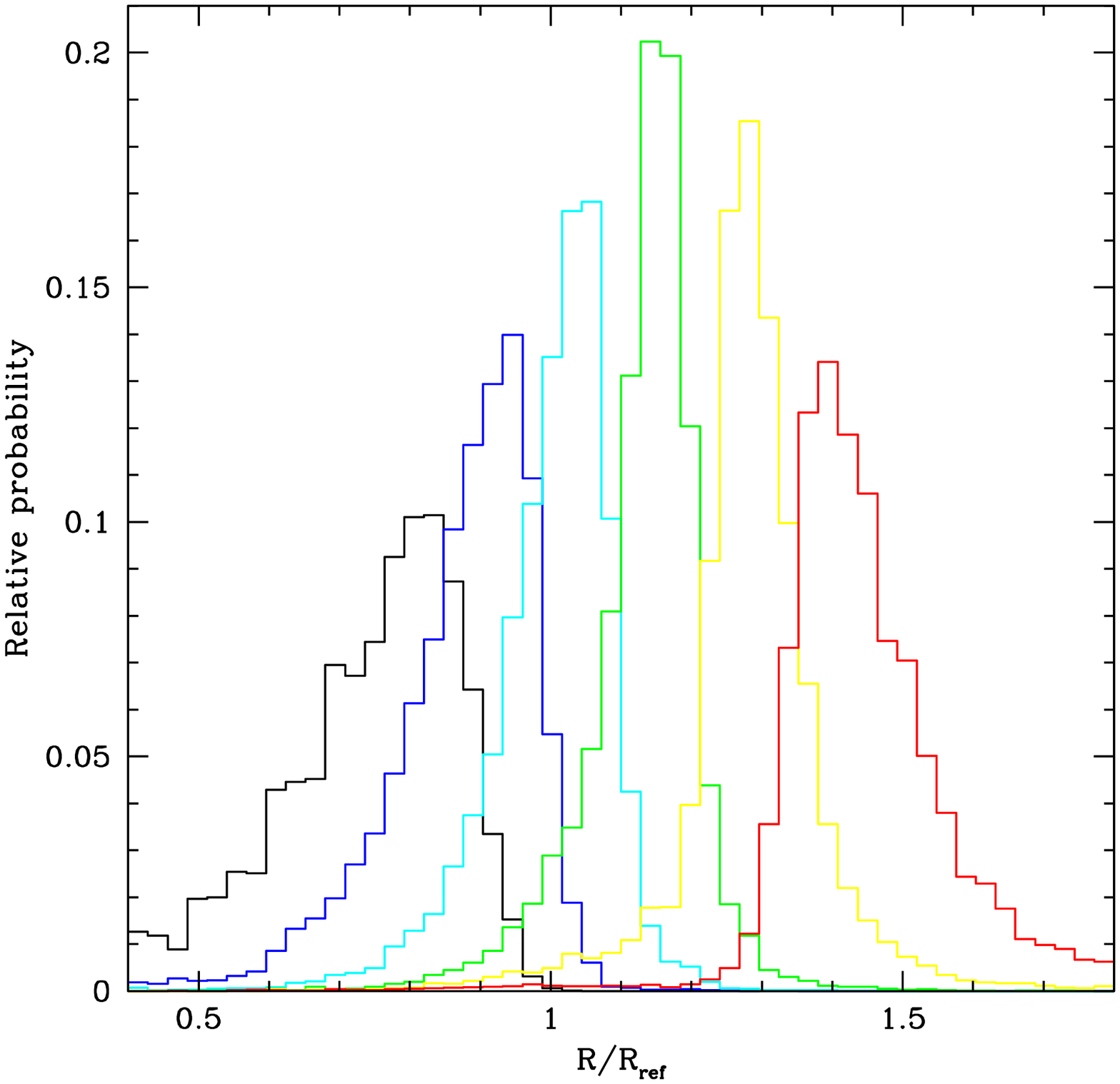}
\caption{\emph{Left :} The final energy profile
obtained from the whole ensemble of source trajectories, including
the prior on the velocity. 
The reference wavelength is $\lambda_{ref}=2125$~\AA\ measured in the source frame.
The error bars give the $1\sigma$ uncertainties. The solid line indicates the linear regression
across the points. The dashed line shows the relation expected for
the standard $R\propto \lambda^{4/3}$ scaling.
\emph{Right :} The corresponding probability distributions of the $R_i$ values
obtained in the  six bands $i=1,2,..., 6$, and using the same color code as
in Fig.~\ref{lightcurves}.}
\label{EP_all}
\end{center}
\end{figure*}

\section{Interpretation of the spectroscopic data}

\subsection{Method}
			
We will now compare the theoretical energy profile with our observations.
In  the procedure described in the  previous sections, we have built a
library of  simulated lightcurves by tracing  a large number  of source
trajectories  through the  magnification  patterns  for various source
sizes $\hat{R}_s$  in order to reproduce the  OGLE V-band photometry $\Delta
m$ of \obj.  We use  this ensemble of  models as an  input to  extend our
analysis to our spectroscopic observations. We consider the integrated
lightcurves in the six color bands defined in Table~\ref{bands} 
as shown in Fig.~\ref{lightcurves}.
For each  band $i$, we have  a lightcurve $\Delta m(\lambda_i,t)$
with 39 epochs.  The wavelength $\lambda_i$ is  the central wavelength of
band $i$ as defined in Table~\ref{bands}.
We then determine which scaled source size  $\hat{R}_i$   best reproduces the
observed lightcurve in band $i$.

Each source  trajectory $j$ in the library is defined  by a set of the
parameters $p_j=(\kappa_j, \gamma_j, \langle M \rangle_j, \hat{R}_{s,j}, 
\hat{V}_j, m_{0,j},\theta_j, \mathbf{x}_{0,A,j}, \mathbf{x}_{0,B,j})$.  
For every trajectory $j$, we  keep all these parameters fixed, except the
source  size $\hat{R}_{s,j}$, that we modify, i.e. we trace the same source trajectory in all
the convolved magnification patterns. By doing this, we get the simulated difference
lightcurve $\Delta m_j^{\prime}(t)$ corresponding to the 45 different 
source sizes defined in Section~\ref{micro_simul}. 
We then
interpolate the magnitude values of  these extracted light curves to
obtain continuous values for the source size.
For every trajectory $j$, this defines a surface 
$\Delta m_j^{\prime}(\hat{R}_s,t)$ in the three-dimensional
space defined by $\hat{R}_s$, $t$ and $\Delta m$.

We use these $\Delta m_j^{\prime}(\hat{R}_s,t)$
to interpret our spectroscopic data
by determining
which scaled source size $\hat{R}_{i}$ best reproduces the observed 
lighcurve $\Delta m(\lambda_i,t)$  in  band $i$.
The size $\hat{R}_{i}$ is the value that  minimizes  the
$\chi_{ij}^2$ between the spectrophotometric observations $\Delta m(\lambda_i,t)$ and
the simulated difference lightcurve $\Delta  m_j^{\prime}(\hat{R}_{i},t)$.  After  this second
fitting   procedure,  we     update    the  relative likelihoods of the lightcurves
$$  
L(D | p_j,i)=\Gamma\left[\frac{n_{dof}-2}{2},\frac{\chi_{OGLE}^2(p_j)}{2}\right]
~\times ~\exp\left(-\frac{\chi_{ij}^2}{2}\right)
$$ 
where  $i$ refers to the considered band.
In the fitting of the spectroscopic observations $\Delta m(\lambda_i,t)$, 
we are comparing similar light curves, where only  $\hat{R}_i$ is changing.
This is a very different situation than when we were fitting the OGLE difference
lightcurve and comparing simulations with very different parameters. 
Hence,   we can  use here
the standard maximum likelihood estimator $\exp(-\chi_{ij}^2/2)$
to compute the likelihoods  $L(D | p_j,i)$.
We run this procedure for all the simulated source trajectories $j$
in the library, and  use these updated likelihoods  to compute the
final probability distribution for the parameters of interest, as described in 
section~\ref{bayes}.


\begin{figure}[t!]
\begin{center}
\includegraphics[width=8.5cm]{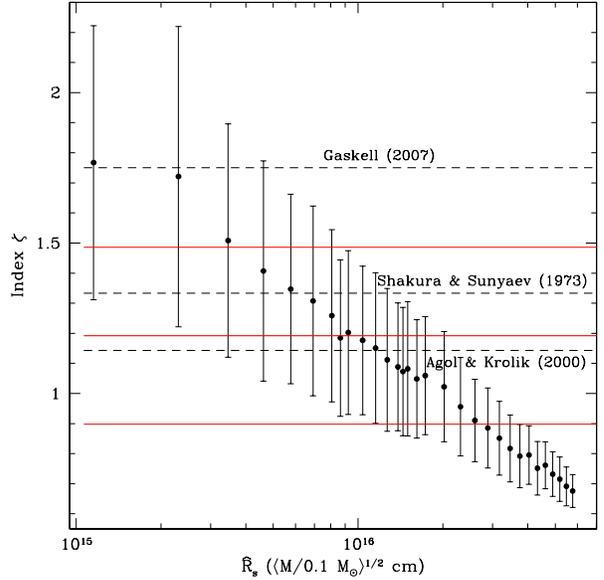}
\caption{The power-law index $\zeta$ of the energy profile $R\propto\lambda^{\zeta}$ 
as a function of the scaled FWHM $\hat{R}_s$ of the source. The points are 
obtained from the subsamples of trajectories with given 
$\hat{R}_s$ fitting the OGLE data and including the prior on the velocity. 
The horizontal solid red lines indicate the index $\zeta$
and its $1\sigma$ deviation obtained by considering the whole library of trajectories 
and including the velocity prior. The dashed lines show the expected
indices for the three indicated accretion-disk models.}
\label{zeta_r}
\end{center}
\end{figure}

\subsection{Results}

For  each  band $i$, we compute the probability distribution  of the
source-size ratio $\hat{R}_i/\hat{R}_{ref}$  
following the Bayesian analysis described in 
Section~\ref{bayes} using
equation~(\ref{prob}).   
The reference radius $\hat{R}_{ref}$ at the reference  
wavelength $\lambda_{ref}$ can  be
chosen arbitrarily,  and we simply  define the  middle band  \#3 as our
reference. 
The scaling between $R_s$ and $\hat{R}_s$ is independent of the
considered wavelength
and $R_i/R_{ref}=\hat{R}_i/\hat{R}_{ref}$. Thus, the ratio $R_i/R_{ref}$
is not expected to depend on the assumed microlens mass $\langle M \rangle$.
We plot the ratio $R_i/R_{ref}$
against the corresponding wavelength ratio $\lambda_i/\lambda_{ref}$, 
and determine the slope of the energy profile by fitting 
a power law 
$$
R\propto \lambda^{\zeta} ~\textrm{.}
$$
We do this for the entire library of source trajectories, but also
individually for every subsample of the library with the same initial 
source size $\hat{R}_s$ used to fit the OGLE data. The resulting index $\zeta$
is plotted as a function of $\hat{R}_s$ in Fig.~\ref{zeta_r}.
We give two examples of energy profiles in Fig.~\ref{R_lambda} derived 
from subsamples of trajectories: one having $\hat{R}_s=0.1\,r_E=10$~pixels and
another with $\hat{R}_s=1\,r_E=100$~pixels. 
The final energy profile obtained from the whole 
sample of source trajectories is shown in Fig.~\ref{EP_all}, given
in Table~\ref{results},  and yields
$\zeta=1.2\pm0.3$ when adding the velocity prior. Without this 
prior we get a flatter energy profile with $\zeta=1.1\pm0.3$.

\begin{table}
\begin{center}
\caption{ \label{results} The relative size $R_{i}/R_{ref}$ obtained
for the six photometric bands, including the velocity prior. 
}
\begin{tabular}{c c c c}
\hline
Band & $\lambda_{c}$~[\AA ]  & $R_{i}/R_{ref}$   \\
\hline      
1 & 1625 & 0.73 $\pm$  0.14 \\
2 & 1875 & 0.87 $\pm$  0.11 \\
3 & 2125 & 1.00 $\pm$  0.09 \\
4 & 2375 & 1.13 $\pm$  0.09 \\
5 & 2625 & 1.29 $\pm$  0.13 \\
6 & 2875 & 1.46 $\pm$  0.18 \\
\hline
\end{tabular}
\end{center}
\end{table}


\section{Discussion}

From the results of our microlensing 
simulations, we can extract several interesting facts.
First, the source FWHM $\hat{R}_s = \left(0.16^{+0.12}_{-0.10} \right) ~r_E$ 
we derive from the fitting of the OGLE data (including the prior on the velocity)
is well compatible with the upper limit of $0.98~r_E$ given by \cite{yonehara01}.
Our result is also   in good agreement  
with the FWHM derived by 
\cite{anguita08}, $\hat{R}_s = (0.06 \pm 0.01) ~r_E$, and  by 
\cite{kochanek04}, $\hat{R}_s = \left(0.20^{+0.19}_{-0.12} \right) ~r_E$.
%
%
%
%
%
%
%
%
%
The differences between these estimates of $\hat{R}_s$ can have two origins.
First, as mentioned by \cite{kochanek04}, the probability distributions 
obtained for the parameters depend on the choice of the period of observations.
If the considered period is very active in terms of microlensing, the simulations
will favor high transverse velocities and/or small source sizes. 
Second, the choice of the velocity prior has a strong effect on the derived source size.  
As we observe in Fig.~\ref{distrib}, there is a strong correlation between
the scaled source size $\hat{R}_s$ and the scaled effective transverse velocity $\hat{V}$. 
Independently of the velocity prior considered, we find that 
the source size is directly proportional to the transverse velocity, 
with the relation 
$\hat{R}_s/\hat{V}=\left( 2.1 \pm 0.6 \right) \times 10^{12}$~cm/(km/s).
This is in excellent agreement with \cite{kochanek04}, who also finds 
$\hat{R}_s/\hat{V}= 2.1 \times 10^{12} $~cm/(km/s), where $R_s$ is expressed as the FWHM.
This defines the time scale of the observed microlensing-induced fluctuations, 
which is given by the half-light radius
divided by the effective transverse velocity, i.e.   
$0.5~\hat{R}_s/\hat{V}= 4.0\pm1.0$~months, and which is 
independent of microlens masses or velocity priors.
%
%
%
%

The strong correlation between $\hat{R}_s$ and $\hat{V}$ \citep{kochanek04}
implies that 
the choice of the velocity prior and the selection of the observation period
can both bias our estimate of the source size.
The microlensing activity during our observations was not as important as during the
periods studied by \cite{kochanek04} and \cite{anguita08}. Furthermore, \cite{anguita08} 
considered a shorter observation period, and used the velocity prior of \cite{gilmerino05}, 
which is more constraining towards high velocities (hence favoring smaller source sizes)
than the one we considered in the present study.
This explains, why our derived $\hat{R}_s$ is slightly larger.
One possible solution to minimize the bias induced by the selection of the observation period
is to consider the longest possible lightcurves, but
this has the evident drawback of dramatically increasing the computing effort, making
the problem rapidly intractable. However, as discussed in the following, the
estimate of the source size has only a limited effect on our determination
of the energy profile.
 
From our library of simulated source trajectories fitting the OGLE data, we extend
our analysis to our spectroscopic data, and determine the energy profile 
$R\propto \lambda^{\zeta}$ of the quasar accretion disk.
When we consider the whole library and include
the prior on the velocity, we get $\zeta=1.2\pm0.3$, which is in good 
agreement with the model of \cite{agol00} and 
with the $\zeta=4/3$ index expected from the standard
accretion disk model. It is less compatible (at $1\sigma$)
with the accretion disk model of \cite{gaskell07}.
Our findings are also in good agreement with \cite{anguita08}, who
obtain $\left(6400~\textrm{\AA} / 4700~\textrm{\AA}\right)^{\zeta}=1.45^{+0.90}_{-0.25}$, which
yields a power-law index $\zeta=1.2^{+2.0}_{-0.6}$.
Without the velocity prior, we get a  lower value
$\zeta=1.1\pm0.3$, which is still compatible with $\zeta=8/7$
and $4/3$, but less  with the model of \cite{gaskell07}. 
This is expected because, without the prior, large sources
gain relatively more importance in the Bayesian analysis, 
and favor flatter energy profile, as shown in Fig.~\ref{zeta_r}. 
However, as discussed earlier, large source sizes  imply
unlikely high transverse velocities, and hence a much more reliable
result is obtained when including the velocity prior. 

We observe that our result for the $R\propto \lambda^{\zeta}$ scaling is not 
very sensitive on the velocity prior with respect to other parameters. For instance,
we know that the source size $\hat{R}_s$ is extremely sensitive to such a prior, and can vary
over orders of magnitudes if the considered transverse
velocity does. On the contrary, the index $\zeta$
varies only by a few percent, even if the considered 
velocities are modified by orders of magnitude. Hence, and independently of the
determination of $\hat{V}$ and $\hat{R}_s$, the chromatic variations 
between two images of a lensed quasar are extremely efficient in constraining  
the relative sizes of different regions of the accretion disk.
This is further confirmed by the fact that \cite{anguita08} obtain a value 
for $\zeta$ which is in good agreement with ours, even though
they derive a slightly smaller source size.

The influence of  the initial value of the parameter
$\hat{R}_s$ (chosen to fit the OGLE data) on the resulting energy profile of
the  accretion disk is obvious in Fig.~\ref{zeta_r}. 
Sources having sizes in the range $0.06 \le \hat{R}_s \le 0.2\, r_E$  
are in good  agreement with the $\zeta=8/7$ and $4/3$ scalings, 
while larger sources give flatter energy profiles, and smaller sources
steeper profiles. The same behavior is observed, when we omit the velocity prior.

\section{Conclusion}

We present the continuation of our spectrophotometric monitoring of \obj\ 
conducted at the Very Large Telescope of the European Southern Observatory,
which extends over more than three years from October 2004 to December 2007.
Our program provides the spectra of the four lensed images at 43 epochs.
Analysis of these data show that the continuum and the broad line
region of the background quasar are microlensed, and that
images A and B are particularly affected. In this paper,
we focus on the observed variations in the continuum of the 
spectra of these two images, and use them to constrain the energy profile
of the quasar.
 
We build microlensing magnification patterns with the inverse 
ray-shooting method \citep{wambsganssPhD,wambsganss90}, and convolve
them with different source sizes. We trace source trajectories through
these convolved patterns and fit the observed OGLE difference lightcurve
between image A and image B. 
Combining these simulations with Bayesian analysis following the method proposed by
\cite{kochanek04}, we infer probability distributions for the  effective transverse 
velocity $V$, and the source size $R_s$ of the quasar. Our results are compatible
with previous studies. 

Besides the OGLE broad-band photometry, we analyze the data
of our spectroscopic monitoring and derived the corresponding difference
lightcurves in six different 
photometric bands. Each band is 250~\AA\ wide and together the bands cover the
wavelength range between 1500 and 3000~\AA\ (measured in the rest frame of the source). 
We compute the difference lightcurves
between images A and B in these six bands, and observe that,
as expected from the microlensing of an accretion disk,
bands at bluer wavelengths exhibit stronger flux variations than bands
at redder wavelengths. 

Based on these chromatic variations and using the ensemble of the 
microlensing simulations that are fitting the 
OGLE data, we determine the relative sizes of the regions of the accretion 
disk emitting in the six photometric bands, i.e.
we derive the energy profile of the accretion disk. We find that this profile
follows a power-law $R\propto \lambda^{\zeta}$ with
$\zeta=1.2\pm0.3$, which is in good agreement with the
standard thin accretion disk model of \cite{shakura73}  ($\zeta=4/3$)
and  with the model of \cite{agol00} ($\zeta=8/7$), where
the disk is powered by the spin of the central black hole. 
Our result is  less compatible with the model 
of \cite{gaskell07}, which predicts a steeper energy profile 
($\zeta=1.75$).
Our result compares well with that obtained by \cite{poindexter08} 
from multi-band photometry of the lensed quasar 
HE~1104$-$1805. They analyzed microlensing lightcurves 
in eleven bands from the optical to the mid-infrared, 
and found that both the size and energy profile 
$\zeta=1.64^{+0.46}_{-0.56}$ (i.e. their $\beta^{-1}$)
of the quasar accretion disk
are consistent with the standard disk model of \cite{shakura73}. 

We observe that the determination of the power law-index is 
almost independent of the velocity prior used, whereas the
determination of the scaled source size $\hat{R}_s$ is directly proportional
to the scaled effective transverse velocity $\hat{V}$. 
This is easily explained by the fact that the scaling between $R_s$ and $\hat{R}_s$ 
is independent of the considered wavelength, and hence 
$R_i/R_{ref}=\hat{R}_i/\hat{R}_{ref}$. 
As a consequence, the determination
of relative source sizes, and hence of the energy profile,
is  not expected to depend on the assumed  microlens
mass $\langle M \rangle$.

Quasar microlensing is hence able to resolve structures
of accretion disks on scales reaching
$0.1$ micro-arcsecond, which
is more than $10,000$ times better than the resolution reached by today's 
best telescopes.
Finally, we should mention several  recent studies that 
compare microlensing  in the X-ray and optical domain, and 
that further demonstrate the efficiency of 
microlensing in probing the inner parts of quasars. 
For instance, \cite{morgan08} analyze the lightcurves of 
the lensed quasar PG~1115$+$080, and find that the effective
radius of the X-ray emission is $1.3^{+1.1}_{-0.5}$~dex smaller than that of
the optical emission, with the X-ray emission
generated near the inner edge of the accretion disk while the 
optical emission comes from scales slightly larger than those 
expected for a standard thin disk. 
\cite{pooley08} observe extreme microlensing-induced (de-) magnification
of the lensed images of PG~1115$+$080, and 
conclude that about 90\% of the matter in the lensing galaxy
is in smoothly distributed (dark) material and only about 
10\% is in compact (stellar) objects.
Another example is given by \cite{chartas08}, who 
combine X-ray and optical data of HE~1104$-$1805, and 
reveal that the X-ray emitting region 
is compact with a half-light radius smaller than 
six gravitational radius, i.e. smaller than $2 \times 10^{15}$~cm,
thus placing significant constraints on active galactic nuclei
models.



\begin{acknowledgements}
This project is partially supported by the Swiss National 
Science Foundation (SNSF).
\end{acknowledgements}


\bibliographystyle{aa} 
\bibliography{0729}

\scriptsize
\begin{flushleft}
\begin{longtable}{cccccccc}
\caption{\label{mag_bands} Continuum integrated AB magnitudes in the six photometric bands defined in Table~\ref{bands}.} \\
\hline 
\hline 
ID & HJD& band 1 (mag) & band 2 (mag) & band 3 (mag) & band 4 (mag) & band 5 (mag) &  band 6 (mag)\\
\endfirsthead
\caption{Continued.} \\
\hline 
\hline
ID & HJD& band 1 (mag) & band 2 (mag) & band 3 (mag) & band 4 (mag) & band 5 (mag) &  band 6 (mag)\\
\hline
\endhead
\hline
\endfoot
\hline  
Image A \\
\hline					       
 1 &3292 &$  17.97 \pm 0.05 $ &$  17.81 \pm 0.05 $ &$  17.67 \pm 0.05$ &$  17.54 \pm 0.06 $ &$  17.43 \pm 0.06 $ & $  17.33 \pm 0.05 $ \\ 
 2 &3324 &$  18.02 \pm 0.05 $ &$  17.87 \pm 0.04 $ &$  17.73 \pm 0.04$ &$  17.61 \pm 0.05 $ &$  17.50 \pm 0.05 $ & $  17.40 \pm 0.04 $ \\ 
 3 &3341 &$  18.02 \pm 0.06 $ &$  17.85 \pm 0.05 $ &$  17.71 \pm 0.05$ &$  17.59 \pm 0.06 $ &$  17.47 \pm 0.06 $ & $  17.37 \pm 0.05 $ \\ 
 4 &3355 &$  17.98 \pm 0.08 $ &$  17.83 \pm 0.08 $ &$  17.69 \pm 0.07$ &$  17.57 \pm 0.08 $ &$  17.47 \pm 0.08 $ & $  17.37 \pm 0.08 $ \\ 
 5 &3502 &$  18.07 \pm 0.04 $ &$  17.88 \pm 0.04 $ &$  17.72 \pm 0.04$ &$  17.57 \pm 0.04 $ &$  17.44 \pm 0.04 $ & $  17.33 \pm 0.04 $ \\ 
 6 &3523 &$  18.16 \pm 0.05 $ &$  17.97 \pm 0.05 $ &$  17.80 \pm 0.04$ &$  17.64 \pm 0.05 $ &$  17.51 \pm 0.05 $ & $  17.39 \pm 0.04 $ \\ 
 7 &3553 &$  18.23 \pm 0.04 $ &$  18.03 \pm 0.04 $ &$  17.85 \pm 0.03$ &$  17.69 \pm 0.04 $ &$  17.55 \pm 0.04 $ & $  17.42 \pm 0.03 $ \\ 
 8 &3566 &$  18.16 \pm 0.06 $ &$  17.96 \pm 0.06 $ &$  17.79 \pm 0.05$ &$  17.63 \pm 0.06 $ &$  17.49 \pm 0.06 $ & $  17.37 \pm 0.05 $ \\ 
 9 &3589 &$  18.24 \pm 0.04 $ &$  18.04 \pm 0.04 $ &$  17.87 \pm 0.03$ &$  17.71 \pm 0.04 $ &$  17.57 \pm 0.04 $ & $  17.45 \pm 0.03 $ \\ 
10 &3598 &$  18.24 \pm 0.06 $ &$  18.03 \pm 0.06 $ &$  17.84 \pm 0.06$ &$  17.68 \pm 0.06 $ &$  17.53 \pm 0.06 $ & $  17.39 \pm 0.06 $ \\ 
11 &3608 &$  18.21 \pm 0.03 $ &$  18.02 \pm 0.03 $ &$  17.86 \pm 0.03$ &$  17.71 \pm 0.04 $ &$  17.58 \pm 0.04 $ & $  17.47 \pm 0.03 $ \\ 
12 &3626 &$  18.21 \pm 0.05 $ &$  18.01 \pm 0.05 $ &$  17.83 \pm 0.04$ &$  17.68 \pm 0.05 $ &$  17.54 \pm 0.05 $ & $  17.42 \pm 0.04 $ \\ 
13 &3641 &$  18.11 \pm 0.05 $ &$  17.90 \pm 0.05 $ &$  17.73 \pm 0.05$ &$  17.57 \pm 0.05 $ &$  17.43 \pm 0.05 $ & $  17.30 \pm 0.05 $ \\ 
14 &3645 &$  18.08 \pm 0.05 $ &$  17.91 \pm 0.05 $ &$  17.77 \pm 0.05$ &$  17.64 \pm 0.05 $ &$  17.52 \pm 0.05 $ & $  17.42 \pm 0.05 $ \\ 
15 &3655 &$  18.10 \pm 0.04 $ &$  17.92 \pm 0.04 $ &$  17.77 \pm 0.03$ &$  17.63 \pm 0.04 $ &$  17.51 \pm 0.04 $ & $  17.39 \pm 0.03 $ \\ 
16 &3665 &$  18.10 \pm 0.04 $ &$  17.92 \pm 0.04 $ &$  17.75 \pm 0.04$ &$  17.61 \pm 0.04 $ &$  17.48 \pm 0.04 $ & $  17.36 \pm 0.04 $ \\ 
17 &3686 &$  17.93 \pm 0.06 $ &$  17.77 \pm 0.06 $ &$  17.63 \pm 0.06$ &$  17.50 \pm 0.07 $ &$  17.39 \pm 0.07 $ & $  17.28 \pm 0.06 $ \\ 
18 &3699 &$  17.94 \pm 0.04 $ &$  17.78 \pm 0.04 $ &$  17.64 \pm 0.04$ &$  17.51 \pm 0.04 $ &$  17.40 \pm 0.04 $ & $  17.30 \pm 0.04 $ \\ 
19 &3711 &$  17.88 \pm 0.07 $ &$  17.71 \pm 0.07 $ &$  17.57 \pm 0.07$ &$  17.44 \pm 0.07 $ &$  17.33 \pm 0.07 $ & $  17.22 \pm 0.07 $ \\ 
20 &3880 &$  17.58 \pm 0.04 $ &$  17.47 \pm 0.05 $ &$  17.36 \pm 0.04$ &$  17.26 \pm 0.05 $ &$  17.17 \pm 0.05 $ & $  17.08 \pm 0.04 $ \\ 
21 &3903 &$  17.50 \pm 0.03 $ &$  17.40 \pm 0.04 $ &$  17.30 \pm 0.04$ &$  17.20 \pm 0.04 $ &$  17.12 \pm 0.04 $ & $  17.04 \pm 0.04 $ \\ 
22 &3907 &$  17.50 \pm 0.03 $ &$  17.39 \pm 0.03 $ &$  17.28 \pm 0.03$ &$  17.19 \pm 0.04 $ &$  17.10 \pm 0.04 $ & $  17.02 \pm 0.03 $ \\ 
23 &3914 &$  17.51 \pm 0.02 $ &$  17.41 \pm 0.03 $ &$  17.30 \pm 0.03$ &$  17.21 \pm 0.03 $ &$  17.12 \pm 0.03 $ & $  17.04 \pm 0.03 $ \\ 
24 &3944 &$  17.53 \pm 0.04 $ &$  17.39 \pm 0.04 $ &$  17.27 \pm 0.04$ &$  17.16 \pm 0.04 $ &$  17.07 \pm 0.04 $ & $  16.98 \pm 0.04 $ \\ 
25 &3951 &$  17.51 \pm 0.03 $ &$  17.39 \pm 0.03 $ &$  17.28 \pm 0.03$ &$  17.18 \pm 0.03 $ &$  17.10 \pm 0.04 $ & $  17.02 \pm 0.03 $ \\ 
26 &4022 &$  17.54 \pm 0.03 $ &$  17.41 \pm 0.04 $ &$  17.29 \pm 0.03$ &$  17.19 \pm 0.04 $ &$  17.09 \pm 0.04 $ & $  17.01 \pm 0.03 $ \\ 
27 &4037 &$  17.53 \pm 0.03 $ &$  17.40 \pm 0.03 $ &$  17.29 \pm 0.03$ &$  17.19 \pm 0.04 $ &$  17.09 \pm 0.04 $ & $  17.01 \pm 0.03 $ \\ 
28 &4050 &$  17.44 \pm 0.04 $ &$  17.33 \pm 0.04 $ &$  17.23 \pm 0.04$ &$  17.14 \pm 0.04 $ &$  17.06 \pm 0.04 $ & $  16.99 \pm 0.04 $ \\ 
29 &4067 &$  17.41 \pm 0.05 $ &$  17.29 \pm 0.05 $ &$  17.19 \pm 0.05$ &$  17.09 \pm 0.05 $ &$  17.01 \pm 0.05 $ & $  16.93 \pm 0.05 $ \\ 
31 &4092 &$  17.40 \pm 0.04 $ &$  17.30 \pm 0.05 $ &$  17.20 \pm 0.04$ &$  17.12 \pm 0.05 $ &$  17.05 \pm 0.05 $ & $  16.99 \pm 0.04 $ \\ 
32 &4292 &$  17.51 \pm 0.03 $ &$  17.36 \pm 0.03 $ &$  17.22 \pm 0.03$ &$  17.10 \pm 0.03 $ &$  16.99 \pm 0.03 $ & $  16.89 \pm 0.03 $ \\ 
33 &4297 &$  17.49 \pm 0.03 $ &$  17.33 \pm 0.03 $ &$  17.19 \pm 0.03$ &$  17.06 \pm 0.03 $ &$  16.95 \pm 0.03 $ & $  16.85 \pm 0.03 $ \\ 
34 &4307 &$  17.42 \pm 0.03 $ &$  17.31 \pm 0.03 $ &$  17.21 \pm 0.03$ &$  17.12 \pm 0.04 $ &$  17.04 \pm 0.04 $ & $  16.97 \pm 0.03 $ \\ 
35 &4316 &$  17.46 \pm 0.03 $ &$  17.34 \pm 0.03 $ &$  17.23 \pm 0.03$ &$  17.13 \pm 0.03 $ &$  17.04 \pm 0.03 $ & $  16.97 \pm 0.03 $ \\ 
36 &4340 &$  17.43 \pm 0.04 $ &$  17.29 \pm 0.04 $ &$  17.17 \pm 0.04$ &$  17.06 \pm 0.04 $ &$  16.97 \pm 0.04 $ & $  16.88 \pm 0.04 $ \\ 
37 &4350 &$  17.44 \pm 0.03 $ &$  17.36 \pm 0.03 $ &$  17.29 \pm 0.03$ &$  17.23 \pm 0.03 $ &$  17.18 \pm 0.03 $ & $  17.13 \pm 0.03 $ \\ 
38 &4364 &$  17.40 \pm 0.03 $ &$  17.28 \pm 0.03 $ &$  17.18 \pm 0.03$ &$  17.09 \pm 0.03 $ &$  17.00 \pm 0.03 $ & $  16.93 \pm 0.03 $ \\ 
39 &4367 &$  17.35 \pm 0.06 $ &$  17.23 \pm 0.06 $ &$  17.11 \pm 0.06$ &$  17.02 \pm 0.07 $ &$  16.93 \pm 0.07 $ & $  16.85 \pm 0.06 $ \\ 
40 &4379 &$  17.39 \pm 0.02 $ &$  17.26 \pm 0.02 $ &$  17.15 \pm 0.02$ &$  17.05 \pm 0.03 $ &$  16.96 \pm 0.03 $ & $  16.87 \pm 0.02 $ \\ 
41 &4384 &$  17.41 \pm 0.03 $ &$  17.29 \pm 0.03 $ &$  17.20 \pm 0.03$ &$  17.11 \pm 0.04 $ &$  17.03 \pm 0.04 $ & $  16.96 \pm 0.03 $ \\ 
42 &4420 &$  17.39 \pm 0.05 $ &$  17.25 \pm 0.05 $ &$  17.13 \pm 0.05$ &$  17.02 \pm 0.05 $ &$  16.93 \pm 0.05 $ & $  16.84 \pm 0.05 $ \\ 
43 &4436 &$  17.39 \pm 0.03 $ &$  17.23 \pm 0.04 $ &$  17.09 \pm 0.03$ &$  16.96 \pm 0.04 $ &$  16.85 \pm 0.04 $ & $  16.75 \pm 0.03 $ \\ 
\hline 
Image B \\
\hline	
 1 &3293 &$  19.08 \pm 0.12 $ &$  18.94 \pm 0.12 $ &$  18.82 \pm 0.11$ &$  18.71 \pm 0.13 $ &$  18.62 \pm 0.13 $ & $  18.53 \pm 0.11 $ \\ 
 2 &3324 &$  19.29 \pm 0.10 $ &$  19.11 \pm 0.09 $ &$  18.94 \pm 0.08$ &$  18.79 \pm 0.09 $ &$  18.66 \pm 0.09 $ & $  18.54 \pm 0.09 $ \\ 
 3 &3341 &$  19.18 \pm 0.14 $ &$  18.99 \pm 0.14 $ &$  18.83 \pm 0.12$ &$  18.69 \pm 0.14 $ &$  18.57 \pm 0.14 $ & $  18.45 \pm 0.13 $ \\ 
 4 &3356 &$  19.17 \pm 0.14 $ &$  19.00 \pm 0.14 $ &$  18.85 \pm 0.13$ &$  18.72 \pm 0.14 $ &$  18.60 \pm 0.14 $ & $  18.49 \pm 0.13 $ \\ 
 5 &3503 &$  18.43 \pm 0.05 $ &$  18.35 \pm 0.05 $ &$  18.28 \pm 0.04$ &$  18.22 \pm 0.05 $ &$  18.17 \pm 0.05 $ & $  18.12 \pm 0.05 $ \\ 
 6 &3523 &$  18.46 \pm 0.06 $ &$  18.36 \pm 0.06 $ &$  18.28 \pm 0.06$ &$  18.21 \pm 0.06 $ &$  18.15 \pm 0.06 $ & $  18.09 \pm 0.06 $ \\ 
 9 &3589 &$  18.76 \pm 0.05 $ &$  18.61 \pm 0.06 $ &$  18.48 \pm 0.05$ &$  18.37 \pm 0.06 $ &$  18.26 \pm 0.06 $ & $  18.17 \pm 0.05 $ \\ 
11 &3608 &$  18.85 \pm 0.05 $ &$  18.68 \pm 0.05 $ &$  18.53 \pm 0.05$ &$  18.40 \pm 0.05 $ &$  18.28 \pm 0.05 $ & $  18.18 \pm 0.05 $ \\ 
12 &3626 &$  18.67 \pm 0.06 $ &$  18.55 \pm 0.06 $ &$  18.44 \pm 0.06$ &$  18.35 \pm 0.07 $ &$  18.26 \pm 0.07 $ & $  18.18 \pm 0.06 $ \\ 
13 &3641 &$  18.61 \pm 0.06 $ &$  18.48 \pm 0.07 $ &$  18.38 \pm 0.06$ &$  18.28 \pm 0.07 $ &$  18.20 \pm 0.07 $ & $  18.12 \pm 0.06 $ \\ 
14 &3645 &$  18.62 \pm 0.07 $ &$  18.49 \pm 0.08 $ &$  18.38 \pm 0.07$ &$  18.28 \pm 0.08 $ &$  18.19 \pm 0.08 $ & $  18.11 \pm 0.07 $ \\ 
15 &3655 &$  18.73 \pm 0.06 $ &$  18.58 \pm 0.06 $ &$  18.45 \pm 0.06$ &$  18.34 \pm 0.06 $ &$  18.24 \pm 0.06 $ & $  18.15 \pm 0.06 $ \\ 
16 &3665 &$  18.72 \pm 0.06 $ &$  18.58 \pm 0.06 $ &$  18.46 \pm 0.06$ &$  18.35 \pm 0.07 $ &$  18.25 \pm 0.07 $ & $  18.16 \pm 0.06 $ \\ 
17 &3686 &$  18.63 \pm 0.10 $ &$  18.50 \pm 0.10 $ &$  18.38 \pm 0.09$ &$  18.28 \pm 0.10 $ &$  18.19 \pm 0.10 $ & $  18.10 \pm 0.09 $ \\ 
18 &3699 &$  18.66 \pm 0.08 $ &$  18.53 \pm 0.08 $ &$  18.41 \pm 0.08$ &$  18.31 \pm 0.09 $ &$  18.22 \pm 0.09 $ & $  18.13 \pm 0.08 $ \\ 
19 &3711 &$  18.56 \pm 0.11 $ &$  18.42 \pm 0.11 $ &$  18.30 \pm 0.10$ &$  18.20 \pm 0.11 $ &$  18.10 \pm 0.11 $ & $  18.01 \pm 0.10 $ \\ 
20 &3880 &$  18.83 \pm 0.12 $ &$  18.74 \pm 0.13 $ &$  18.59 \pm 0.12$ &$  18.47 \pm 0.14 $ &$  18.35 \pm 0.14 $ & $  18.25 \pm 0.12 $ \\ 
21 &3903 &$  18.89 \pm 0.07 $ &$  18.77 \pm 0.07 $ &$  18.60 \pm 0.07$ &$  18.45 \pm 0.08 $ &$  18.31 \pm 0.08 $ & $  18.19 \pm 0.07 $ \\ 
22 &3907 &$  18.83 \pm 0.06 $ &$  18.71 \pm 0.07 $ &$  18.53 \pm 0.07$ &$  18.38 \pm 0.07 $ &$  18.25 \pm 0.07 $ & $  18.12 \pm 0.07 $ \\ 
23 &3914 &$  18.86 \pm 0.05 $ &$  18.74 \pm 0.06 $ &$  18.57 \pm 0.05$ &$  18.42 \pm 0.06 $ &$  18.29 \pm 0.06 $ & $  18.17 \pm 0.05 $ \\ 
24 &3944 &$  18.76 \pm 0.07 $ &$  18.60 \pm 0.07 $ &$  18.47 \pm 0.06$ &$  18.35 \pm 0.07 $ &$  18.24 \pm 0.07 $ & $  18.15 \pm 0.06 $ \\ 
25 &3951 &$  18.72 \pm 0.06 $ &$  18.56 \pm 0.06 $ &$  18.43 \pm 0.06$ &$  18.31 \pm 0.06 $ &$  18.20 \pm 0.06 $ & $  18.10 \pm 0.06 $ \\ 
26 &4022 &$  18.76 \pm 0.06 $ &$  18.61 \pm 0.06 $ &$  18.48 \pm 0.05$ &$  18.36 \pm 0.06 $ &$  18.25 \pm 0.06 $ & $  18.15 \pm 0.05 $ \\ 
27 &4037 &$  18.69 \pm 0.05 $ &$  18.54 \pm 0.05 $ &$  18.41 \pm 0.05$ &$  18.29 \pm 0.05 $ &$  18.19 \pm 0.05 $ & $  18.09 \pm 0.05 $ \\ 
28 &4050 &$  18.63 \pm 0.09 $ &$  18.49 \pm 0.09 $ &$  18.36 \pm 0.08$ &$  18.25 \pm 0.09 $ &$  18.15 \pm 0.09 $ & $  18.06 \pm 0.08 $ \\ 
29 &4067 &$  18.59 \pm 0.12 $ &$  18.46 \pm 0.12 $ &$  18.34 \pm 0.12$ &$  18.23 \pm 0.12 $ &$  18.13 \pm 0.12 $ & $  18.05 \pm 0.12 $ \\ 
30 &4089 &$  18.57 \pm 0.14 $ &$  18.45 \pm 0.14 $ &$  18.34 \pm 0.13$ &$  18.25 \pm 0.14 $ &$  18.16 \pm 0.14 $ & $  18.08 \pm 0.13 $ \\ 
31 &4093 &$  18.60 \pm 0.10 $ &$  18.46 \pm 0.10 $ &$  18.34 \pm 0.09$ &$  18.23 \pm 0.10 $ &$  18.13 \pm 0.10 $ & $  18.05 \pm 0.09 $ \\ 
32 &4292 &$  18.66 \pm 0.06 $ &$  18.49 \pm 0.07 $ &$  18.34 \pm 0.06$ &$  18.21 \pm 0.07 $ &$  18.09 \pm 0.06 $ & $  17.98 \pm 0.06 $ \\ 
33 &4297 &$  18.66 \pm 0.07 $ &$  18.49 \pm 0.07 $ &$  18.35 \pm 0.06$ &$  18.22 \pm 0.07 $ &$  18.11 \pm 0.07 $ & $  18.00 \pm 0.06 $ \\ 
34 &4307 &$  18.54 \pm 0.09 $ &$  18.40 \pm 0.09 $ &$  18.28 \pm 0.08$ &$  18.17 \pm 0.09 $ &$  18.08 \pm 0.09 $ & $  17.99 \pm 0.08 $ \\ 
35 &4316 &$  18.66 \pm 0.08 $ &$  18.50 \pm 0.09 $ &$  18.36 \pm 0.08$ &$  18.24 \pm 0.09 $ &$  18.12 \pm 0.09 $ & $  18.02 \pm 0.08 $ \\ 
36 &4340 &$  18.69 \pm 0.13 $ &$  18.51 \pm 0.13 $ &$  18.35 \pm 0.12$ &$  18.20 \pm 0.13 $ &$  18.07 \pm 0.13 $ & $  17.96 \pm 0.12 $ \\ 
37 &4350 &$  18.59 \pm 0.08 $ &$  18.46 \pm 0.08 $ &$  18.34 \pm 0.08$ &$  18.23 \pm 0.08 $ &$  18.14 \pm 0.08 $ & $  18.05 \pm 0.08 $ \\ 
38 &4364 &$  18.60 \pm 0.09 $ &$  18.44 \pm 0.09 $ &$  18.31 \pm 0.09$ &$  18.20 \pm 0.10 $ &$  18.09 \pm 0.09 $ & $  18.00 \pm 0.09 $ \\ 
39 &4367 &$  18.59 \pm 0.16 $ &$  18.46 \pm 0.17 $ &$  18.34 \pm 0.15$ &$  18.24 \pm 0.17 $ &$  18.15 \pm 0.17 $ & $  18.07 \pm 0.16 $ \\ 
40 &4379 &$  18.62 \pm 0.05 $ &$  18.45 \pm 0.06 $ &$  18.30 \pm 0.05$ &$  18.17 \pm 0.06 $ &$  18.05 \pm 0.06 $ & $  17.94 \pm 0.05 $ \\ 
41 &4384 &$  18.68 \pm 0.06 $ &$  18.51 \pm 0.07 $ &$  18.36 \pm 0.06$ &$  18.24 \pm 0.07 $ &$  18.12 \pm 0.07 $ & $  18.01 \pm 0.06 $ \\ 
42 &4421 &$  18.72 \pm 0.09 $ &$  18.54 \pm 0.09 $ &$  18.39 \pm 0.08$ &$  18.25 \pm 0.09 $ &$  18.13 \pm 0.09 $ & $  18.02 \pm 0.08 $ \\ 
43 &4436 &$  18.60 \pm 0.07 $ &$  18.43 \pm 0.07 $ &$  18.28 \pm 0.06$ &$  18.14 \pm 0.07 $ &$  18.02 \pm 0.07 $ & $  17.91 \pm 0.06 $ \\  
\hline 
Image C \\
\hline	
 1 &3293 &$  19.08 \pm 0.11 $ &$  18.92 \pm 0.12 $ &$  18.78 \pm 0.11$ &$  18.65 \pm 0.13 $ &$  18.54 \pm 0.13 $ & $  18.44 \pm 0.11 $ \\ 
 2 &3324 &$  19.32 \pm 0.09 $ &$  19.09 \pm 0.08 $ &$  18.90 \pm 0.08$ &$  18.73 \pm 0.09 $ &$  18.57 \pm 0.09 $ & $  18.43 \pm 0.08 $ \\ 
 3 &3341 &$  19.22 \pm 0.14 $ &$  19.01 \pm 0.14 $ &$  18.82 \pm 0.13$ &$  18.65 \pm 0.14 $ &$  18.50 \pm 0.14 $ & $  18.37 \pm 0.12 $ \\ 
 4 &3356 &$  19.44 \pm 0.16 $ &$  19.22 \pm 0.16 $ &$  19.02 \pm 0.16$ &$  18.85 \pm 0.18 $ &$  18.69 \pm 0.18 $ & $  18.55 \pm 0.15 $ \\ 
 5 &3503 &$  19.28 \pm 0.08 $ &$  19.09 \pm 0.08 $ &$  18.93 \pm 0.08$ &$  18.79 \pm 0.09 $ &$  18.66 \pm 0.08 $ & $  18.54 \pm 0.08 $ \\ 
 6 &3523 &$  19.29 \pm 0.10 $ &$  19.09 \pm 0.10 $ &$  18.91 \pm 0.10$ &$  18.76 \pm 0.11 $ &$  18.62 \pm 0.10 $ & $  18.50 \pm 0.09 $ \\ 
 9 &3589 &$  19.46 \pm 0.09 $ &$  19.23 \pm 0.09 $ &$  19.02 \pm 0.08$ &$  18.83 \pm 0.09 $ &$  18.67 \pm 0.09 $ & $  18.52 \pm 0.08 $ \\ 
11 &3608 &$  19.59 \pm 0.09 $ &$  19.32 \pm 0.09 $ &$  19.10 \pm 0.08$ &$  18.89 \pm 0.09 $ &$  18.71 \pm 0.08 $ & $  18.54 \pm 0.08 $ \\ 
12 &3626 &$  19.38 \pm 0.10 $ &$  19.17 \pm 0.11 $ &$  18.99 \pm 0.10$ &$  18.83 \pm 0.11 $ &$  18.68 \pm 0.11 $ & $  18.55 \pm 0.10 $ \\ 
13 &3641 &$  19.21 \pm 0.10 $ &$  19.02 \pm 0.10 $ &$  18.85 \pm 0.10$ &$  18.70 \pm 0.11 $ &$  18.56 \pm 0.10 $ & $  18.44 \pm 0.09 $ \\ 
14 &3645 &$  19.25 \pm 0.12 $ &$  19.04 \pm 0.12 $ &$  18.86 \pm 0.12$ &$  18.70 \pm 0.13 $ &$  18.56 \pm 0.13 $ & $  18.43 \pm 0.11 $ \\ 
15 &3655 &$  19.44 \pm 0.09 $ &$  19.20 \pm 0.10 $ &$  19.00 \pm 0.09$ &$  18.82 \pm 0.10 $ &$  18.65 \pm 0.10 $ & $  18.50 \pm 0.09 $ \\ 
16 &3665 &$  19.41 \pm 0.10 $ &$  19.19 \pm 0.10 $ &$  18.99 \pm 0.09$ &$  18.81 \pm 0.10 $ &$  18.66 \pm 0.10 $ & $  18.51 \pm 0.09 $ \\ 
17 &3686 &$  19.19 \pm 0.15 $ &$  18.98 \pm 0.15 $ &$  18.80 \pm 0.14$ &$  18.63 \pm 0.15 $ &$  18.48 \pm 0.15 $ & $  18.35 \pm 0.13 $ \\ 
18 &3699 &$  19.24 \pm 0.12 $ &$  19.03 \pm 0.12 $ &$  18.84 \pm 0.11$ &$  18.68 \pm 0.12 $ &$  18.53 \pm 0.12 $ & $  18.40 \pm 0.11 $ \\ 
19 &3711 &$  18.94 \pm 0.14 $ &$  18.74 \pm 0.14 $ &$  18.57 \pm 0.14$ &$  18.42 \pm 0.15 $ &$  18.28 \pm 0.15 $ & $  18.16 \pm 0.13 $ \\ 
20 &3880 &$  18.86 \pm 0.18 $ &$  18.76 \pm 0.20 $ &$  18.61 \pm 0.20$ &$  18.47 \pm 0.22 $ &$  18.35 \pm 0.21 $ & $  18.24 \pm 0.19 $ \\ 
21 &3903 &$  19.19 \pm 0.12 $ &$  19.05 \pm 0.14 $ &$  18.84 \pm 0.13$ &$  18.67 \pm 0.14 $ &$  18.51 \pm 0.14 $ & $  18.36 \pm 0.13 $ \\ 
22 &3907 &$  19.12 \pm 0.13 $ &$  18.98 \pm 0.14 $ &$  18.77 \pm 0.14$ &$  18.59 \pm 0.15 $ &$  18.43 \pm 0.15 $ & $  18.29 \pm 0.13 $ \\ 
23 &3914 &$  19.15 \pm 0.10 $ &$  19.01 \pm 0.12 $ &$  18.81 \pm 0.11$ &$  18.63 \pm 0.12 $ &$  18.47 \pm 0.12 $ & $  18.32 \pm 0.11 $ \\ 
24 &3944 &$  19.01 \pm 0.08 $ &$  18.83 \pm 0.09 $ &$  18.67 \pm 0.08$ &$  18.52 \pm 0.09 $ &$  18.39 \pm 0.09 $ & $  18.28 \pm 0.08 $ \\ 
25 &3951 &$  18.99 \pm 0.07 $ &$  18.80 \pm 0.08 $ &$  18.64 \pm 0.07$ &$  18.50 \pm 0.08 $ &$  18.37 \pm 0.07 $ & $  18.25 \pm 0.07 $ \\ 
26 &4022 &$  19.14 \pm 0.11 $ &$  18.94 \pm 0.12 $ &$  18.76 \pm 0.11$ &$  18.61 \pm 0.12 $ &$  18.46 \pm 0.12 $ & $  18.34 \pm 0.10 $ \\ 
27 &4037 &$  19.07 \pm 0.11 $ &$  18.87 \pm 0.11 $ &$  18.70 \pm 0.10$ &$  18.55 \pm 0.11 $ &$  18.41 \pm 0.11 $ & $  18.29 \pm 0.10 $ \\ 
28 &4050 &$  18.83 \pm 0.11 $ &$  18.65 \pm 0.11 $ &$  18.50 \pm 0.11$ &$  18.36 \pm 0.12 $ &$  18.24 \pm 0.11 $ & $  18.12 \pm 0.11 $ \\ 
29 &4067 &$  18.78 \pm 0.19 $ &$  18.61 \pm 0.20 $ &$  18.47 \pm 0.19$ &$  18.34 \pm 0.20 $ &$  18.22 \pm 0.20 $ & $  18.11 \pm 0.19 $ \\ 
30 &4089 &$  18.80 \pm 0.21 $ &$  18.62 \pm 0.23 $ &$  18.47 \pm 0.22$ &$  18.33 \pm 0.23 $ &$  18.21 \pm 0.23 $ & $  18.10 \pm 0.20 $ \\ 
31 &4093 &$  18.92 \pm 0.20 $ &$  18.72 \pm 0.20 $ &$  18.55 \pm 0.19$ &$  18.40 \pm 0.21 $ &$  18.27 \pm 0.20 $ & $  18.15 \pm 0.18 $ \\ 
32 &4292 &$  18.75 \pm 0.06 $ &$  18.57 \pm 0.06 $ &$  18.42 \pm 0.06$ &$  18.28 \pm 0.06 $ &$  18.16 \pm 0.06 $ & $  18.05 \pm 0.06 $ \\ 
33 &4297 &$  18.73 \pm 0.06 $ &$  18.57 \pm 0.07 $ &$  18.42 \pm 0.06$ &$  18.29 \pm 0.07 $ &$  18.18 \pm 0.07 $ & $  18.07 \pm 0.06 $ \\ 
34 &4307 &$  18.57 \pm 0.08 $ &$  18.43 \pm 0.09 $ &$  18.31 \pm 0.08$ &$  18.20 \pm 0.09 $ &$  18.11 \pm 0.09 $ & $  18.02 \pm 0.08 $ \\ 
35 &4316 &$  18.71 \pm 0.08 $ &$  18.55 \pm 0.09 $ &$  18.42 \pm 0.08$ &$  18.29 \pm 0.09 $ &$  18.18 \pm 0.09 $ & $  18.08 \pm 0.08 $ \\ 
36 &4340 &$  18.79 \pm 0.12 $ &$  18.59 \pm 0.13 $ &$  18.42 \pm 0.12$ &$  18.26 \pm 0.13 $ &$  18.12 \pm 0.13 $ & $  17.99 \pm 0.12 $ \\ 
37 &4350 &$  18.79 \pm 0.09 $ &$  18.63 \pm 0.09 $ &$  18.50 \pm 0.08$ &$  18.38 \pm 0.09 $ &$  18.27 \pm 0.09 $ & $  18.18 \pm 0.08 $ \\ 
38 &4364 &$  18.80 \pm 0.09 $ &$  18.62 \pm 0.10 $ &$  18.46 \pm 0.09$ &$  18.32 \pm 0.10 $ &$  18.19 \pm 0.10 $ & $  18.08 \pm 0.09 $ \\ 
39 &4367 &$  18.50 \pm 0.14 $ &$  18.37 \pm 0.15 $ &$  18.26 \pm 0.14$ &$  18.17 \pm 0.16 $ &$  18.08 \pm 0.16 $ & $  18.00 \pm 0.14 $ \\ 
40 &4379 &$  18.98 \pm 0.07 $ &$  18.77 \pm 0.08 $ &$  18.58 \pm 0.07$ &$  18.41 \pm 0.08 $ &$  18.26 \pm 0.07 $ & $  18.12 \pm 0.07 $ \\ 
41 &4384 &$  19.03 \pm 0.09 $ &$  18.82 \pm 0.10 $ &$  18.64 \pm 0.09$ &$  18.48 \pm 0.10 $ &$  18.34 \pm 0.10 $ & $  18.21 \pm 0.09 $ \\ 
42 &4421 &$  19.18 \pm 0.13 $ &$  18.94 \pm 0.13 $ &$  18.74 \pm 0.12$ &$  18.55 \pm 0.14 $ &$  18.39 \pm 0.13 $ & $  18.24 \pm 0.12 $ \\ 
43 &4436 &$  19.06 \pm 0.09 $ &$  18.83 \pm 0.10 $ &$  18.63 \pm 0.10$ &$  18.45 \pm 0.10 $ &$  18.29 \pm 0.10 $ & $  18.14 \pm 0.09 $ \\ 
\hline 	
Image D \\
\hline
 1 &3292 &$  19.11 \pm 0.14 $ &$  18.91 \pm 0.14 $ &$  18.73 \pm 0.14$ &$  18.57 \pm 0.15 $ &$  18.42 \pm 0.15 $ & $  18.29 \pm 0.13 $ \\ 
 2 &3324 &$  19.23 \pm 0.12 $ &$  19.03 \pm 0.12 $ &$  18.86 \pm 0.12$ &$  18.71 \pm 0.13 $ &$  18.58 \pm 0.13 $ & $  18.45 \pm 0.12 $ \\ 
 3 &3341 &$  19.18 \pm 0.14 $ &$  18.98 \pm 0.14 $ &$  18.79 \pm 0.14$ &$  18.63 \pm 0.15 $ &$  18.49 \pm 0.15 $ & $  18.36 \pm 0.13 $ \\ 
 4 &3355 &$  19.04 \pm 0.18 $ &$  18.84 \pm 0.19 $ &$  18.67 \pm 0.18$ &$  18.51 \pm 0.21 $ &$  18.38 \pm 0.20 $ & $  18.25 \pm 0.18 $ \\ 
 5 &3502 &$  19.25 \pm 0.11 $ &$  19.02 \pm 0.11 $ &$  18.81 \pm 0.10$ &$  18.63 \pm 0.11 $ &$  18.47 \pm 0.11 $ & $  18.32 \pm 0.10 $ \\ 
 6 &3523 &$  19.29 \pm 0.11 $ &$  19.08 \pm 0.11 $ &$  18.89 \pm 0.11$ &$  18.73 \pm 0.12 $ &$  18.58 \pm 0.12 $ & $  18.45 \pm 0.11 $ \\ 
 7 &3553 &$  19.42 \pm 0.10 $ &$  19.20 \pm 0.10 $ &$  19.01 \pm 0.09$ &$  18.84 \pm 0.11 $ &$  18.68 \pm 0.10 $ & $  18.55 \pm 0.09 $ \\ 
 8 &3566 &$  19.26 \pm 0.14 $ &$  19.01 \pm 0.14 $ &$  18.80 \pm 0.13$ &$  18.61 \pm 0.15 $ &$  18.44 \pm 0.14 $ & $  18.28 \pm 0.13 $ \\ 
 9 &3589 &$  19.44 \pm 0.09 $ &$  19.23 \pm 0.09 $ &$  19.04 \pm 0.09$ &$  18.87 \pm 0.10 $ &$  18.73 \pm 0.10 $ & $  18.59 \pm 0.09 $ \\ 
10 &3598 &$  19.35 \pm 0.15 $ &$  19.10 \pm 0.15 $ &$  18.87 \pm 0.14$ &$  18.68 \pm 0.15 $ &$  18.50 \pm 0.15 $ & $  18.34 \pm 0.13 $ \\ 
11 &3608 &$  19.43 \pm 0.09 $ &$  19.22 \pm 0.09 $ &$  19.05 \pm 0.08$ &$  18.89 \pm 0.10 $ &$  18.75 \pm 0.10 $ & $  18.62 \pm 0.09 $ \\ 
12 &3626 &$  19.43 \pm 0.12 $ &$  19.20 \pm 0.12 $ &$  18.99 \pm 0.11$ &$  18.81 \pm 0.12 $ &$  18.65 \pm 0.12 $ & $  18.50 \pm 0.11 $ \\ 
13 &3641 &$  19.27 \pm 0.12 $ &$  19.02 \pm 0.12 $ &$  18.79 \pm 0.11$ &$  18.60 \pm 0.13 $ &$  18.42 \pm 0.13 $ & $  18.26 \pm 0.11 $ \\ 
14 &3645 &$  19.26 \pm 0.12 $ &$  19.03 \pm 0.12 $ &$  18.83 \pm 0.11$ &$  18.65 \pm 0.13 $ &$  18.50 \pm 0.13 $ & $  18.35 \pm 0.11 $ \\ 
15 &3655 &$  19.33 \pm 0.10 $ &$  19.13 \pm 0.10 $ &$  18.96 \pm 0.09$ &$  18.81 \pm 0.11 $ &$  18.67 \pm 0.11 $ & $  18.54 \pm 0.10 $ \\ 
16 &3665 &$  19.33 \pm 0.11 $ &$  19.10 \pm 0.11 $ &$  18.91 \pm 0.10$ &$  18.73 \pm 0.12 $ &$  18.58 \pm 0.11 $ & $  18.43 \pm 0.10 $ \\ 
17 &3686 &$  19.03 \pm 0.15 $ &$  18.82 \pm 0.15 $ &$  18.64 \pm 0.15$ &$  18.48 \pm 0.16 $ &$  18.33 \pm 0.16 $ & $  18.20 \pm 0.14 $ \\ 
18 &3699 &$  19.14 \pm 0.11 $ &$  18.94 \pm 0.11 $ &$  18.76 \pm 0.11$ &$  18.60 \pm 0.12 $ &$  18.46 \pm 0.12 $ & $  18.33 \pm 0.10 $ \\ 
19 &3711 &$  18.95 \pm 0.17 $ &$  18.75 \pm 0.18 $ &$  18.58 \pm 0.17$ &$  18.42 \pm 0.19 $ &$  18.28 \pm 0.18 $ & $  18.16 \pm 0.16 $ \\ 
20 &3880 &$  19.00 \pm 0.11 $ &$  18.83 \pm 0.13 $ &$  18.66 \pm 0.12$ &$  18.51 \pm 0.14 $ &$  18.37 \pm 0.13 $ & $  18.25 \pm 0.12 $ \\ 
21 &3903 &$  19.08 \pm 0.10 $ &$  18.93 \pm 0.12 $ &$  18.77 \pm 0.11$ &$  18.63 \pm 0.13 $ &$  18.50 \pm 0.13 $ & $  18.39 \pm 0.11 $ \\ 
22 &3907 &$  19.07 \pm 0.08 $ &$  18.92 \pm 0.09 $ &$  18.75 \pm 0.09$ &$  18.61 \pm 0.10 $ &$  18.48 \pm 0.10 $ & $  18.36 \pm 0.09 $ \\ 
23 &3914 &$  19.13 \pm 0.07 $ &$  18.98 \pm 0.08 $ &$  18.83 \pm 0.07$ &$  18.70 \pm 0.08 $ &$  18.58 \pm 0.08 $ & $  18.47 \pm 0.08 $ \\ 
24 &3944 &$  19.08 \pm 0.12 $ &$  18.87 \pm 0.12 $ &$  18.69 \pm 0.12$ &$  18.52 \pm 0.13 $ &$  18.38 \pm 0.13 $ & $  18.24 \pm 0.11 $ \\ 
25 &3951 &$  19.04 \pm 0.11 $ &$  18.85 \pm 0.10 $ &$  18.68 \pm 0.10$ &$  18.53 \pm 0.11 $ &$  18.40 \pm 0.11 $ & $  18.28 \pm 0.10 $ \\ 
26 &4022 &$  19.06 \pm 0.07 $ &$  18.88 \pm 0.08 $ &$  18.72 \pm 0.07$ &$  18.58 \pm 0.08 $ &$  18.46 \pm 0.08 $ & $  18.34 \pm 0.07 $ \\ 
27 &4037 &$  19.03 \pm 0.07 $ &$  18.85 \pm 0.07 $ &$  18.70 \pm 0.07$ &$  18.56 \pm 0.08 $ &$  18.44 \pm 0.08 $ & $  18.32 \pm 0.07 $ \\ 
28 &4050 &$  18.93 \pm 0.12 $ &$  18.75 \pm 0.12 $ &$  18.59 \pm 0.12$ &$  18.45 \pm 0.13 $ &$  18.32 \pm 0.13 $ & $  18.20 \pm 0.12 $ \\ 
29 &4067 &$  18.84 \pm 0.11 $ &$  18.65 \pm 0.11 $ &$  18.49 \pm 0.10$ &$  18.35 \pm 0.12 $ &$  18.22 \pm 0.12 $ & $  18.11 \pm 0.10 $ \\ 
31 &4092 &$  18.87 \pm 0.09 $ &$  18.70 \pm 0.09 $ &$  18.55 \pm 0.09$ &$  18.42 \pm 0.10 $ &$  18.30 \pm 0.10 $ & $  18.19 \pm 0.09 $ \\ 
32 &4292 &$  19.00 \pm 0.09 $ &$  18.78 \pm 0.09 $ &$  18.58 \pm 0.09$ &$  18.41 \pm 0.10 $ &$  18.25 \pm 0.09 $ & $  18.11 \pm 0.08 $ \\ 
33 &4297 &$  18.95 \pm 0.08 $ &$  18.73 \pm 0.08 $ &$  18.53 \pm 0.08$ &$  18.36 \pm 0.09 $ &$  18.21 \pm 0.08 $ & $  18.07 \pm 0.08 $ \\ 
34 &4307 &$  18.81 \pm 0.11 $ &$  18.63 \pm 0.12 $ &$  18.48 \pm 0.12$ &$  18.34 \pm 0.13 $ &$  18.22 \pm 0.13 $ & $  18.11 \pm 0.11 $ \\ 
35 &4316 &$  18.85 \pm 0.11 $ &$  18.69 \pm 0.12 $ &$  18.54 \pm 0.11$ &$  18.41 \pm 0.12 $ &$  18.30 \pm 0.12 $ & $  18.20 \pm 0.11 $ \\ 
36 &4340 &$  18.73 \pm 0.13 $ &$  18.55 \pm 0.13 $ &$  18.39 \pm 0.13$ &$  18.24 \pm 0.14 $ &$  18.12 \pm 0.14 $ & $  18.00 \pm 0.12 $ \\ 
37 &4350 &$  18.86 \pm 0.11 $ &$  18.75 \pm 0.12 $ &$  18.65 \pm 0.12$ &$  18.57 \pm 0.14 $ &$  18.50 \pm 0.14 $ & $  18.43 \pm 0.12 $ \\ 
38 &4364 &$  18.76 \pm 0.10 $ &$  18.59 \pm 0.11 $ &$  18.44 \pm 0.10$ &$  18.31 \pm 0.11 $ &$  18.19 \pm 0.11 $ & $  18.09 \pm 0.10 $ \\ 
39 &4367 &$  18.59 \pm 0.18 $ &$  18.41 \pm 0.19 $ &$  18.26 \pm 0.18$ &$  18.12 \pm 0.20 $ &$  17.99 \pm 0.20 $ & $  17.88 \pm 0.18 $ \\ 
40 &4379 &$  18.70 \pm 0.07 $ &$  18.52 \pm 0.08 $ &$  18.36 \pm 0.08$ &$  18.22 \pm 0.08 $ &$  18.09 \pm 0.08 $ & $  17.97 \pm 0.07 $ \\ 
41 &4384 &$  18.74 \pm 0.09 $ &$  18.57 \pm 0.10 $ &$  18.42 \pm 0.10$ &$  18.29 \pm 0.11 $ &$  18.18 \pm 0.11 $ & $  18.07 \pm 0.09 $ \\ 
42 &4420 &$  18.61 \pm 0.12 $ &$  18.43 \pm 0.13 $ &$  18.27 \pm 0.13$ &$  18.13 \pm 0.14 $ &$  18.01 \pm 0.14 $ & $  17.89 \pm 0.13 $ \\ 
43 &4436 &$  18.59 \pm 0.09 $ &$  18.39 \pm 0.10 $ &$  18.21 \pm 0.09$ &$  18.05 \pm 0.10 $ &$  17.91 \pm 0.10 $ & $  17.78 \pm 0.09 $ \\ 
\end{longtable}  
\end{flushleft}

\normalsize

\end{document}